\documentclass[%
 aip,
 amsmath,amssymb,
 reprint,%
]{revtex4-1}

\usepackage{graphicx}
\usepackage{dcolumn}
\usepackage{bm}

\usepackage[utf8]{inputenc}
\usepackage[T1]{fontenc}
\usepackage{mathptmx}
\usepackage{etoolbox}
\usepackage{bigints}
\usepackage[dvipsnames]{xcolor}
\usepackage{comment} 
\usepackage{multirow}

\makeatletter
\def\@email#1#2{%
 \endgroup
 \patchcmd{\titleblock@produce}
  {\frontmatter@RRAPformat}
  {\frontmatter@RRAPformat{\produce@RRAP{*#1\href{mailto:#2}{#2}}}\frontmatter@RRAPformat}
  {}{}
}%
\makeatother
\begin{document}

\preprint{AIP/123-QED}

\title[Dissociation line of H$_{2}$ hydrates]{Dissociation line and driving force for nucleation of the multiple occupied hydrogen hydrate from computer simulation}
\author{Miguel J. Torrej\'on}
\affiliation{Laboratorio de Simulaci\'on Molecular y Qu\'imica Computacional, CIQSO-Centro de Investigaci\'on en Qu\'imica Sostenible and Departamento de Ciencias Integradas, Universidad de Huelva, 21006 Huelva Spain}

\author{S. Blazquez}
\affiliation{Dpto. Qu\'{\i}mica F\'{\i}sica I, Fac. Ciencias Qu\'{\i}micas, Universidad Complutense de Madrid, 28040 Madrid, Spain}

\author{Jes\'us Algaba}
\affiliation{Laboratorio de Simulaci\'on Molecular y Qu\'imica Computacional, CIQSO-Centro de Investigaci\'on en Qu\'imica Sostenible and Departamento de Ciencias Integradas, Universidad de Huelva, 21006 Huelva Spain}

\author{M. M. Conde$^{*}$}
\affiliation{Departamento de Ingeniería Química Industrial y del Medio Ambiente, Escuela Técnica Superior de Ingenieros Industriales, Universidad Politécnica de Madrid,
28006, Madrid, Spain.
}

\author{Felipe J. Blas*}
\affiliation{Laboratorio de Simulaci\'on Molecular y Qu\'imica Computacional, CIQSO-Centro de Investigaci\'on en Qu\'imica Sostenible and Departamento de Ciencias Integradas, Universidad de Huelva, 21006 Huelva Spain}

\begin{abstract}

In this work, we determine the dissociation temperature of the hydrogen (H$_2$) hydrate by computer simulation using two different methods. In both cases, the molecules of water and H$_2$ are modeled using the TIP4P/Ice and a modified version of the Silvera and Goldman models respectively, and the Berthelot combining rule for the cross water-H$_2$ interactions has been modified. The first method used in this work is the solubility method which consists in determining the solubility of H$_2$ in an aqueous phase when in contact with a H$_2$ hydrate (H--L$_{\text{w}}$) phase and when in contact with a pure H$_2$ phase (L$_{\text{w}}$--L$_{\text{H}_2}$) at different temperatures. At a given pressure value, both solubility curves intersect at the temperature ($T_3$) at which the three phases coexist in equilibrium. Following this approach, we determine the dissociation temperature of the H$_2$ hydrate at $185\,\text{MPa}$ finding a good agreement with the data previously reported in the literature. We also analyze the effect of the multiple occupancy of the D, or small, and H, or large, cages of the sII hydrate structure. We conclude that the $T_3$ value is barely affected by the occupancy of the H$_2$ hydrate at $185\,\text{MPa}$. From the analysis of the solubility curves and performing extra bulk simulations of the three phases involved in the equilibrium, we also determine the driving force for nucleation ($\Delta\mu^{\text{EC}}_{N}$) at $185\,\text{MPa}$ as a function of the supercooling degree and the H$_2$ hydrate occupancy. We determine that, thermodynamically, the most favored occupancy of the H$_2$ hydrate consists of 1 H$_2$ molecule in the D cages and 3 in the H cages (named as 1-3 occupancy). We also conclude that the double occupancy of the small D cages is not favored due to the $\Delta\mu^{\text{EC}}_{N}$ values obtained for this occupancy are the most positive ones. The second approach used in this work is the direct coexistence technique using an initial H$_2$ hydrate phase with the 1-3 occupancy. We also propose a new modification of the Berthelot combining rule to improve the predictions of the $T_3$ values. Following this method, we determine the $T_3$ at 100, 185, and $300\,\text{MPa}$ finding and excellent agreement with the experimental data.  

\end{abstract}

\maketitle
$^*$Corresponding authors: felipe@uhu.es and maria.mconde@upm.es\\
\maketitle

%

\section{Introduction}

Clathrates are non-stoichiometric inclusion crystalline compounds consisting of a network of hydrogen-bonded molecules (host) forming cages in which small molecules (guest) such as hydrogen (H$_2$), nitrogen (N$_2$), carbon dioxide (CO$_2$), and methane (CH$_4$), among many others, can be encapsulated under the appropriate thermodynamic conditions.~\cite{Sloan2008a,Ripmeester2022a} When water acts as the host molecule, these structures are specifically referred to as clathrate hydrates or simply hydrates. Among the many hydrates applications such as CO$_2$ capture~\cite{ma2016review,C5CP07202F,dashti2015recent,cannone2021review,duc2007co2,choi2022effective,lee2014quantitative,mi2024molecular} or N$_2$ recovery from industrial emissions,~\cite{Yi2019a,hassanpouryouzband2018co2} it is interesting to remark their energetic applications. In nature, vast deposits of natural gas exist as hydrates. According to the last estimations, conventional natural gas deposits correspond to 20\% of the total natural gas present in nature, while the remaining 80\% is trapped as hydrate on the seabeds and permafrost areas.~\cite{Chong2016a,Zheng2020a,Bourry2007a}  Thus, methane hydrate represents an interesting alternative as a source of energy.

Although the use of hydrates as a source of natural gas would help to alleviate the global energy crisis that the world is facing nowadays, it is also necessary to mitigate anthropogenic carbon emissions in order to palliate climate change. However, reducing carbon emissions while ensuring the energy needs of a growing population is not an easy task. In this regard, the use of hydrates of H$_{2}$, in combination with hydrate promoters, as strategic materials for gas transport and storage, is one of the most significant and promising applications of hydrates in the environmental, energetic, and economic context. This represents a potential alternative to the metal hydrides that are currently in use. However, the use of metal hydrides as H$_2$ storage media has not yet been implemented due to a lack of information regarding the thermodynamic and kinetic properties of these compounds.~\cite{Tsimpanogiannis2017a,Brumby2019a,Yi2019a,Michalis2022a}  The utilization of H$_2$ hydrates in this context would result in a reduction in raw material costs while maintaining a comparable volumetric storage capacity. In order to achieve this, it is necessary a deep understanding of the phase equilibria and the kinetics of the formation and growth of these hydrates, with special emphasis on which are the factors that rule the occupancy of these hydrates.~\cite{Mao2002a,Belosudov2016a,Dyadin1999a,Dyadin1999b,Grim2012a,Mao2004a,Katsumasa2007a,Papadimitriou2016a,Liu2017a,Brumby2019a,Michalis2022a,Belosludov2009a,Alavi2005a,Papadimitriou2008a,Papadimitriou2008b,Chun2008a,Papadimitriou2017a,brumby2016cage}

At high pressures and low temperatures, H$_2$ hydrates crystallize in the so-called sII structure with $136$ water molecules distributed in $16$ D (pentagonal dodecahedron or $5^{12}$) cages and $8$ H (hexakaidecahedron or $5^{12}6^{4}$) cages. The D, or small, cages are better stabilized by small molecules such as H$_2$ or N$_2$ while the H, or large, cages are better stabilized by larger molecules such as propane. However, it is interesting to remark that the large H cages can be also stabilized by the multiple occupancy of small molecules such as H$_2$ or N$_2$.~\cite{Sloan2008a,Ripmeester2022a} From an experimental point of view, the analysis of hydrate occupancy is a challenging endeavor, primarily due to the extreme conditions required for their formation and equilibrium stability. These conditions include high pressure and low temperature, which present significant difficulties in producing homogeneous samples for analysis. Furthermore, the experimental measurement of equilibrium properties, such as the lattice constants and cage occupancies, is not a straightforward process.~\cite{Dyadin1999a,Dyadin1999b,Grim2012a,Mao2002a,Mao2004a,Efimchenko2009a} In a series of recent works, Zhang \emph{et al.}~\cite{Zhang2023b,Zhang2023c,Zhang2024a} provide a detailed analysis of the H$_2$ hydrate formation in the presence of several hydrate promoters (tetrahydrofuran, $_\text{L}$-val, and 1,3-dioxolane). They studied the formation kinetics of the H$_2$ hydrate under the presence of different thermodynamic and kinetic promoters as well as a detailed analysis of how the hydrate promoter concentration affects the H$_2$ hydrate formation. They study the occupancy of both types of sII hydrate cages by using Raman spectroscopy, which is a special interesting topic due to the possibility of using hydrates to store H$_2$, concluding that H$_2$ molecules solely occupy the small D cages of the sII hydrates with thermodynamic hydrate promoters occupying the large H cages. At this point is interesting to remark that although tetrahydrofuran is a very special thermodynamic hydrate promoter since it is able to form a hydrate by itself,~\cite{Asadi2019a,Sun2017a,Suzuki2011a,Sabase2009a,Strauch2018a,Ganji2006a,Andersson1996a,Chong2016a,Kumar2010a,Makino2005a,Algaba2024c,Torrejon2024a} the 1,3-dioxolane is less toxic\cite{Torre2015a} and, hence, more friendly from an environmental point of view, which is a prerequisite for large-scale application of hydrate-based H$_2$ storage technology.~\cite{Zhang2024a}

In order to use hydrates as H$_2$ storage media, it is necessary to obtain a deep understanding of how the thermodynamic conditions affect the occupancy of these compounds, since an increment of the occupancy implies an improvement of the storage capacity. In this respect, theoretical and simulation approaches provide an interesting molecular perspective to study the occupancy of these hydrates. Although the occupancy of H$_2$ hydrates has been the subject of several studies, there is still an open debate about the amount of H$_2$ that can be encapsulated inside the small, D, and large, H, cages of the sII hydrate structure. From an experimental point of view, Mao \emph{et al.}~\cite{Mao2002a,Mao2004a} state that the double occupancy of both types of cage is possible. This has been corroborated by Belosudov \emph{et al.}~\cite{Belosudov2016a} from theoretical calculations and by Liu \emph{et al.}~\cite{Liu2017a} from \emph{ab initio} computations. However, from simulations, there is still controversy over the H$_2$ hydrate occupancy. Alavi \emph{et al.}~\cite{Alavi2005a} state that, at low pressures (below $2.5\,\text{kbar}$), the most stable occupancy of the sII H$_2$ hydrate consists of 4 H$_2$ molecules in the H cages while the D cages remain single occupied. They also remark that the double occupancy of the D cages provokes an increase of the structure energy and tetragonal distortions of the hydrate unit cell. The same conclusion was found by Papadimitriou \emph{et al.}~\cite{Papadimitriou2008a} at pressures between 380 and $450\,\text{MPa}$. They also affirm that the most stable occupancy consists of 4 and 1 H$_2$ molecules in the H and D cages respectively. They also study the effect of lattice constant on the storage capacity of hydrogen hydrates\cite{Papadimitriou2016a} and the impact of different force fields on the H$_2$ hydrate storage predictions.~\cite{Papadimitriou2017a} Also in the works of Katsumasa \emph{et al.}~\cite{Katsumasa2007a} and Chun and Lee\cite{Chun2008a} it is shown that H cages can be multiply occupied while D cages remain single occupied in most cases. Contrary to the previous simulation studies, Brumby \emph{et al.}~\cite{Brumby2019a} performed a detailed analysis of the H$_2$ hydrate occupancy from Gibbs
 ensemble Monte Carlo simulations. They conclude that the occupancy of D cages is not limited to only single occupancy at pressures below $400\,\text{MPa}$, although only a small percentage of them in the hydrate structure were doubly occupied.

In this work, we study the dissociation temperature of the sII H$_2$ hydrate at $185\,\text{MPa}$ as a function of the multiple occupancy of both D and H cages. At the dissociation temperature ($T_3$), the system under study presents a three-phase equilibrium. The three phases involved are a H$_2$ hydrate phase, an aqueous phase with the corresponding equilibrium H$_2$ solubility, and a pure H$_2$ phase. The dissociation temperature at $185\,\text{MPa}$ and different H$_2$ hydrate occupancies is obtained by molecular dynamic simulations and using the solubility method.~\cite{Tanaka2018a,Grabowska2022a, Algaba2023a,Algaba2023b,Torrejon2024b} We use the TIP4P/Ice\cite{Abascal2005b} and a modified version of the Silvera and Goldman\cite{Alavi2005a,Silvera1978a} models to describe the molecules of water and H$_2$ respectively. We analyze the effect of the occupancy of the large H cages from single to quadruple occupancy and the occupancy of the small D cages from single to double occupancy on the dissociation temperature. We also obtain the driving force for nucleation, $\Delta\mu_{\text{N}}$, at $185\,\text{MPa}$ as a function of the H$_2$ hydrate occupancy and the supercooling degree. Taking into account the dissociation temperature and the driving force for nucleation results obtained for the different occupancies, we study the dissociation temperature of the most favored one (3 and 1 H$_2$ molecules in the H and D cages respectively) from 100 to $300\,\text{MPa}$. Finally, we propose a new modification of the Berthelot combining rule to improve the simulation results and predict accurately the experimental data.~\cite{Dyadin1999a,Efimchenko2009a} It is important to remark that, experimentally, hydrate phase diagrams are studied by performing experiments at different thermodynamic conditions and monitoring when a phase transition takes place.~\cite{Dyadin1999a,Efimchenko2009a} Notice that this implies the crystallization of the H$_2$ hydrate from an H$_2$ aqueous solution. Homogeneous nucleation at the thermodynamic equilibrium conditions is a rare event that only can be studied from simulation using special and computationally expensive techniques, which hinders to employ the same specific conditions and procedures as experiments. However, the same information can be obtained by simpler and cheaper simulation methodologies such as the direct coexistence technique\cite{Conde2010a,Conde2013a,Michalis2015a,Miguez2015a,Costandy2015a,Perez-Rodriguez2017a,Yi2019a,Michalis2022a,Algaba2024c,Fernandez-Fernandez2019a,Blazquez2023b,Borrero2025a} and the solubility method.\cite{Tanaka2018a,Grabowska2022a, Algaba2023a,Algaba2023b,Torrejon2024b} Notice that, for a given pressure, there is only a possible $T_3$ value which should be independent of the specific conditions at which experiments and simulations are performed.

The organization of this paper is as follows: In Sec.~II, we describe the models, simulation details, and methodology used in this work. The results obtained, as well as their discussion, are described in Sec.~III. Finally, conclusions are presented in Sec.~IV.

\section{Models, simulation details, and methodology}

In this work, all molecular dynamic simulations have been carried out using the GROMACS~\cite{VanDerSpoel2005a} (2016.5 double--precision version) software package. Water molecules are modeled using the widely-known TIP4P/Ice model.~\cite{Abascal2005b} Following the work of Alavi \emph{et al.,}~\cite{Alavi2005a} we employ a modified version of the Silvera and Goldman\cite{Silvera1978a} (SG) H$_2$ model based on the well depth and potential minimum of the
original SG isotropic isolated pair potential for gas-phase hydrogen. This modified version simplifies the original SG interaction potential and makes it more suitable for high-performance simulations.~\cite{Michalis2022a} In this model, the H$_2$ molecule is described by two hydrogen atoms linked by a rigid bond of $0.7414\text{\AA}$ with a Lennard-Jones (LJ) interactive site in the molecule's center of mass. Also, positive charges are placed on the hydrogen centers while the negative charge is placed in the molecule's center of mass together with the LJ interactive site. A summary of the molecular model details of both compounds is shown in Table \ref{Model}.

In all cases the non-bonded cross interactions between unlike groups are calculated using the Lorentz-Berthelot combining rule. However, as in the original work of Michalis \emph{et al.},~\cite{Michalis2022a} we modify the Berthelot combination rule between the oxygen group from the water molecule and the LJ interactive site from the H$_2$ molecule. This modification improves the predictions of the H$_2$ solubility in water and the three-phase coexistence temperature ($T_3$) determination in the $273.15-323.15\,\text{K}$ and $100-300\,\text{MPa}$ range of temperatures and pressures respectively. In particular, the Berthelot combination rule is modified by a $\chi$ factor whose value is a function of the temperature:

\begin{equation}
    \epsilon_{\text{O-MH$_2$}}=\chi(\epsilon_{\text{OO}}\,\epsilon_\text{{MH$_2$-MH$_2$}})^{1/2}
    \label{unlike}
\end{equation}

\begin{equation}
    \chi=a_1\, \text{exp}\Biggl(-\dfrac{T}{a_2}\Biggr)+a_0+b_0
    \label{chi}
\end{equation}

\noindent where $\chi$ is the Berthelot modifier factor, $\epsilon_{\text{OO}}$ and $\epsilon_\text{{MH$_2$-MH$_2$}}$ are the LJ well-depth associated with the oxygen atom from the water molecule and the center of mass of the H$_2$ molecule respectively. The $a_0$, $a_1$, and $a_2$ parameters are taken from the work of Michalis \emph{et al.}~\cite{Michalis2022a} (we refer the lector to the original work for further details) and the $b_0$ parameter value is proposed in this work. Particularly, the model proposed by these authors corresponds values $a_{0}=0.649194$, $a_{1}=0.12085$, and $a_{2}=-186.29458\,\text{K}$. Notice that when $b_0=0$, Eq.~\eqref{chi} is identical as that proposed by Michalis \emph{et al.}~\cite{Michalis2022a}

Although the original expression for the $\chi$ factor proposed for Michalis \emph{et al.}~\cite{Michalis2022a} provides an excellent agreement between the experimental and the simulated solubility of H$_2$ in water, and also improves the prediction of the three-phase coexistence temperature ($T_3$) in the $100-300\,\text{MPa}$ range of pressure, the $T_3$ values obtained from simulation still underestimate the experimental $T_3$ value (see Section III. D for further details). Hence, in this work, we propose a new $\chi$ expression based on the original one proposed for Michalis \emph{et al.}~\cite{Michalis2022a}. We introduce the $b_0$ parameter in Eq. (\ref{chi}) in order to improve the $T_3$ value obtained from molecular dynamic simulations. In this work, $b_0$ has a value of 0.1. The addition of this factor provides an excellent prediction of the H$_2$ hydrate $T_3$ value but slightly overestimate the H$_2$ solubility in water. This $b_0$ value was obtained by determining the predicted $T_3$ value using different $\chi$ values. In all cases, we used the initial value obtained from the original $\chi$ expression, and $T_{3}$ was evaluated at $100$, $185$, and $300\,\text{MPa}$ by systematically varying $\chi$ in increments of $0.1$. We determine that just a 0.1 increment is enough to provide an accurate description of the $T_3$ value since an additional increment of the $\chi$ factor ($b_0=0.2$) overestimates it. A detailed analysis of the solubility of H$_2$ in water using the TIP4P/Ice and SG models for the water and H$_2$ molecules as a function of the $\chi$ factor can be found in the original work of Michalis \emph{et al.}~\cite{Michalis2022a}

\begin{table}
\caption{Non-bonded interaction parameters and geometry details of TIP4P/Ice water~\cite{Abascal2005b} and modified SG H$_2$~\cite{Alavi2005a,Michalis2022a,Luis2018a} molecular models employed in this work.}
\centering
\begin{tabular}{lccccc}
\hline\hline
Atom &$\sigma(\text{\AA})$ & $\varepsilon/k_B(\text{K})$ & $q$(e) & \multicolumn{2}{c}{Geometry details} \\
\hline
\multicolumn{6}{c}{Water (TIP4P/Ice)} \\
\hline
O            & 3.1668 & 106.1 & - & $d_\text{OH}$ (\AA) & 0.9572\\
H & - & - & 0.5897 & H--O--H (º) & 104.5\\
M  & - & - & -1.1794 & $d_\text{OM}$ (\AA) & 0.1577\\
\hline
\multicolumn{6}{c}{H$_2$ (SG)} \\
\hline
H & 0 & 0 & 0.4932 & \multirow{2}{*}{$d_\text{HH}$ (\AA)} & \multirow{2}{*}{0.7414}\\
Mass center & 3.038 & 34.302 & -0.9864 \\
\hline
\hline
\end{tabular}
\label{Model}
\end{table}

In this work, the Verlet leapfrog\cite{Cuendet2007a} algorithm with a time step of $2\,\text{fs}$ is used to solve the motion equations of Newton. Also, we use the Nosé-Hoover thermostat,~\cite{Nose1984a} with a time constant of $2\,\text{ps}$, and the Parrinello-Rahman barostat,~\cite{Parrinello1981a} with a time constant of $2\,\text{ps}$ and a $5\times 10^{-5}$ compressibility value, to ensure that simulations are performed at constant temperate and pressure. Following the original work of Michalis \emph{et al.},~\cite{Michalis2022a} we use a cut-off value of $1.1\,\text{nm}$ for the Coulombic and dispersive interactions. We don't use long-range corrections for the dispersive LJ interactions but particle-mesh Ewald (PME)~\cite{Essmann1995a} corrections are used for the Coulombic potential.

In order to determine the three-phase coexistence temperature, we employ two different methods. First, we employ the solubility method~\cite{Tanaka2018a,Grabowska2022a, Algaba2023a,Algaba2023b,Torrejon2024b} and the original modification of the $\chi$ factor proposed by Michalis \emph{et al.}~\cite{Michalis2022a} (i.e., $b_0$=0). Following the procedure used in our previous works~\cite{Grabowska2022a, Algaba2023a,Algaba2023b,Torrejon2024b}, the $T_3$ value is determined by studying the solubility of H$_2$ in an aqueous solution phase when it is in contact via a planar interface with an initial pure H$_2$ phase (L$_{\text{w}}$--L$_{\text{H}_{2}}$ two-phase equilibria) and when in contact with a hydrate phase (H--L$_{\text{w}}$ two-phase equilibria). The number of molecules used in each case is specified in their corresponding sections. According to the solubility method, the $T_3$ value is determined by representing the solubility values obtained from both equilibria at a constant pressure and as a function of the temperature. The temperature value at which both solubility curves intersect is the temperature ($T_3$) at which the three phases coexist in equilibrium. The solubility of H$_{2}$ in the aqueous phase when it is in contact with the H$_{2}$-rich liquid phase decreases with temperature, as it happens with the solubility of most gases in water; in
contrast, the solubility of H$_{2}$ when it is contact with the hydrate phase increases with temperature. Note that this behavior with temperature is similar to that observed when a solid is dissolved in water. In the intersection of both curves, the aqueous phase has reached the same thermodynamic equilibrium state (monitoring by the H$_2$ solubility value) when in contact with a H$_2$ hydrate phase and when in contact with a pure liquid H$_2$ phase separately, i.e. the temperature at which
both curves cross is the temperature at which the liquid water, hydrate,
and H$_2$ liquid phases are in equilibrium, which represents the so-called
$T_3$ at the corresponding pressure. In this work, the $T_3$ value is determined using the solubility method and the $\chi$ factor proposed by Michalis \emph{et al.}~\cite{Michalis2022a} at $P=185\,\text{MPa}$. We also analyze the effect of the H$_2$ hydrate occupancy on the $T_3$ value. The sII unit cell hydrate structure is built up by 16 D (small) and 8 H (large) hydrate cages. In this work, we calculated the $T_3$ at $185\,\text{MPa}$ when the D, or small, cages are singly occupied and the H, or large, cages are occupied by 1, 2, 3, and 4 H$_2$ molecules (called from now on 1--1, 1--2, 1--3, and 1--4 occupancies). Also, we analyze the effect of double occupancy in the small D cages when 2 and 4 H$_2$ molecules occupy the small D and large H cages repectively (2--2 and 2--4 occupancies). Unfortunately, although we generate and equilibrate a bulk 2--4 H$_2$ hydrate phase, it becomes unstable when it is put in contact with an aqueous phase. Therefore, we conclude that the 2--4 occupancy is not stable and can not be simulated under the thermodynamic conditions considered in this work. 

The L$_{\text{w}}$-L$_{\text{H}_{2}}$ equilibria simulations are performed in the $NP_{z}T$ ensemble i.e. only the $P_z$ component of the pressure tensor, which is perpendicular to the L$_{\text{w}}$--L$_{\text{H}_{2}}$ planar interface, is fixed by the barostat. The H--L$_{\text{w}}$ equilibria simulations are carried out in the anisotropic $NPT$ ensemble i.e. each side of the simulation box is allowed to fluctuate independently to keep the pressure constant and to avoid any stress from the solid hydrate structure. Also, we perform extra bulk simulations, in the isotropic $NPT$ ensemble,  of pure water, pure H$_2$, and H$_2$ hydrate phases with different occupancy levels to determine the driving force for nucleation as a function of the H$_2$ hydrate occupancy and the supercooling degree. The water and H$_2$ pure bulk phases are built up by 1000 molecules of water and H$_2$ respectively, and the bulk H$_2$ hydrate phase is obtained by replicating the unit cell twice in each space direction taking into account the corresponding H$_2$ occupancy.

The second method used in this work is the direct-coexistence technique.~\cite{Ladd1977a,Vega2008a,Conde2010a,Conde2013a,Michalis2015a,Miguez2015a,Costandy2015a,Perez-Rodriguez2017a,Yi2019a,Michalis2022a,Fernandez-Fernandez2019a,Blazquez2023b,Zhang2022a,Zhang2023a,Fernandez-Fernandez2024a,Algaba2024a,Algaba2024b,Algaba2024c,Blazquez2024a} Following this method, the three phases involved in the equilibrium (H-L$_{\text{w}}$-L$_{\text{H}_{2}}$) are placed together in the same simulation box. By fixing the pressure and varying the temperature, it is possible to analyze the behavior of the three-phase system and determine the $T_3$ value. If the fixed temperature is below the $T_3$, at a given pressure value, the aqueous L$_{\text{w}}$ phase becomes unstable and the hydrate phase grows until extinguishing the L$_{\text{w}}$ or the L$_{\text{H}_{2}}$ phase depending of the number of molecules of water and guest in both phases. Contrary, if the fixed temperature is above the $T_3$ value, the hydrate phase becomes unstable and it melts, obtaining a L$_{\text{w}}$-L$_{\text{H}_{2}}$ equilibria. In this work, the $T_3$ value is obtained at three different pressures (100, 185, and $300\,\text{MPa}$), using an initial hydrate phase seed with a 1-3 occupancy and using the modified $\chi$ factor expression proposed in this work (Eq. (\ref{chi}) with $b_0=0.1$). As in the study of the H--L$_{\text{w}}$ equilibria, simulations are performed in the anisotropic $NPT$ ensemble to avoid any stress from the solid hydrate structure.

\section{Results}

In this section, we present a detailed analysis of the H$_2$ hydrate dissociation temperature. First, we focus on determining the $T_3$ value at $185\,\text{MPa}$ as a function of the H$_2$ hydrate occupancy using the solubility method and a modification of the Berthelot combining rule already presented in the literature by Michalis \emph{et al.}~\cite{Michalis2022a} We also study the driving force for nucleation of the H$_2$ hydrate as a function of the supercooling degree for the five different occupancies studied in this work at $185\,\text{MPa}$. Finally, we propose a modification of the $\chi$ factor expression proposed by Michalis \emph{et al.}~\cite{Michalis2022a} and we determine the $T_3$ at 100, 185, and $300\,\text{MPa}$ using the direct coexistence technique. The new expression for the $\chi$ factor proposed in this work provides the same $T_3$ values from simulations as those reported from experiments in the literature between the error bars. At this point, is important to remark that the $\chi$ original factor expression was proposed to match the simulated and experimental solubility of H$_2$ in an aqueous phase. This resulted in an improvement of the $T_3$ prediction using the classical Lorentz-Berthelot combining rule, but the $T_3$ values obtained with this modification of the Berthelot combining rule still underestimate slightly the experimental $T_3$ value. On the other hand, the new expression for the $\chi$ factor proposed in this work provides an excellent prediction of the $T_3$ values but slightly overestimates the H$_2$ solubility in water.

\subsection{Solubility method}

In order to determine the $T_3$ value at $185\,\text{MPa}$ through the solubility method, first it is necessary to obtain the solubility of H$_2$ in water when an aqueous phase (L$_{\text{w}}$) is in contact with a pure H$_2$ phase (L$_{\text{H}_{2}}$) and when in contact with a H$_2$ hydrate phase (H). First, we focus on the L$_{\text{w}}$--L$_{\text{H}_{2}}$ equilibria and then in the H--L$_{\text{w}}$ equilibria.

\subsubsection{L$_{\text{w}}$--L$_{\text{H}_{2}}$ equilibria: Solubility of H$_2$ in liquid water and interfacial tension}

We study the L$_{\text{w}}$--L$_{\text{H}_{2}}$ equilibria behavior by running $NP_zT$ simulations at $185\,\text{MPa}$ and 230, 240, 250, 260, 270, and $280\,\text{K}$. The initial simulation box is built up by a pure water phase with 2800 water molecules and a pure H$_2$ phase with 1400 H$_2$ molecules. Both phases are put in contact via a planar interface along the $z$-direction. The $xy$ interfacial area remains constant along the whole simulation since the values of $L=Lx=Ly$ are fixed at $3.8\,\text{nm}$. Finally, the average $L_z$ side of the simulation box is $\approx9.8\,\text{nm}$ and both phases approximately fill half of the simulation box along the $z$-axis direction. Due to the low solubility of H$_2$ in water, simulations are run for $800\,\text{ns}$. The first $400\,\text{ns}$ are taken as the equilibration period, and the last $400\,\text{ns}$ as the production period.

\begin{figure}
\includegraphics[width=\columnwidth]{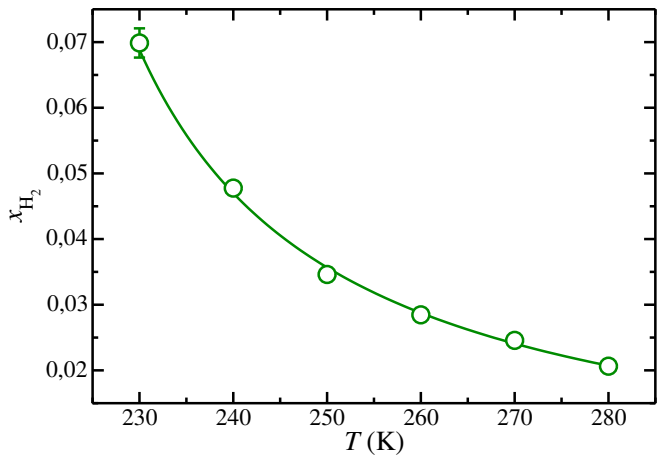}\\
\caption{H$_2$ solubility in the aqueous solution phase as a function of the temperature at $185\,\text{MPa}$. The green open-circle symbols represent the solubility values obtained from the L$_{\text{w}}$--L$_{\text{H}_{2}}$ $NP_zT$ molecular dynamic simulations performed in this work. The green curve is included as a guide to the eye.}
\label{figure1}
\end{figure}

The H$_2$ solubility values in the aqueous phase are calculated from the analysis of the density profiles. These are obtained by dividing the simulation box into 200 slabs and assigning the center of mass of each molecule to the corresponding slab. Finally, the mass density profile is obtained by multiplying the molar density profile of each compound by its corresponding molar mass. The solubility of H$_2$ in the aqueous phase at each temperature is obtained by averaging for each component the values of the density profiles using each slab belonging to the aqueous phase. The error bars of the densities of water and H$_2$ in the aqueous phase at each temperature are obtained as the average standard deviation. The H$_2$ solubility error bars are obtained by propagating the density errors obtained for both components. Calculations are performed far enough from the interface to avoid the effect of the interface on the average H$_2$ solubility determination. Figure \ref{figure1} shows the H$_2$ solubility values in the aqueous phase obtained in this work at $185\,\text{MPa}$ and several temperatures. The results obtained in this work are in good agreement with those reported by Michalis \emph{et al.}~\cite{Michalis2022a} As can be seen in Fig. \ref{figure1}, the solubility of H$_2$ in the aqueous phase decreases when the temperature is increased. This is the expected behavior of the solubility of a gas in water with the temperature. Notice that, when the temperature is increased,  the H$_2$ solubility varies more steeply at the lowest temperature values studied in this work.

\begin{figure}
\includegraphics[width=\columnwidth]{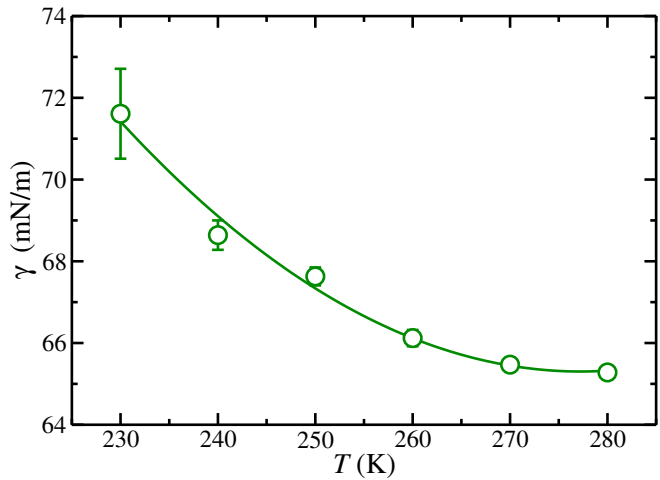}\\
\caption{L$_{\text{w}}$--L$_{\text{H}_{2}}$ interfacial tension, $\gamma$, as a function of the temperature at $185\,\text{MPa}$. The results obtained in this work from molecular dynamic $NP_{z}T$ simulations, and its error bars, are represented as open green circles. The green curve is included as a guide to the eyes.}
\label{figure2}
\end{figure}

From the analysis of the L$_{\text{w}}$--L$_{\text{H}_{2}}$ it is possible to determine the interfacial tension between the aqueous and the H$_2$ phases from the diagonal components of the pressure tensor.~\cite{Rowlinson1982b,deMiguel2006c,deMiguel2006b} As in the case of the density profiles, the interfacial tension at each temperature is determined from the last $400\,\text{ns}$ of the L$_{\text{w}}$--L$_{\text{H}_{2}}$ molecular dynamic simulations. In order to obtain an estimation of the errors, the $400\,\text{ns}$ of the production period are divided into 10 blocks of $40\,\text{ns}$. The final value of the interfacial tension is obtained by averaging the value of each block. Finally, uncertainties are estimated as the standard deviation of the average.~\cite{Flyvbjerg1989a} As it can be observed in Fig \ref{figure2}, the L$_{\text{w}}$--L$_{\text{H}_{2}}$ interfacial tension decreases steeply when the temperature is increased from 230 to $260\,\text{K}$ until reaches an almost plateau value at 270 and $280\,\text{K}$. Similar behavior was found in a previous work\cite{Algaba2023b} when the L$_{\text{w}}$--L$_{\text{N}_{2}}$ interfacial tension was analyzed at 1000 and $1500\,\text{bar}$ from 245 to $300\,\text{K}$. Unfortunately, as far as the authors know there are no experimental data available in the literature to compare with the results obtained in this work. 

\subsubsection{H--L$_{\text{w}}$ equilibria: Solubility of H$_2$ in liquid water when in contact with H$_2$ hydrate.}

Following a similar approach as in the L$_{\text{w}}$--L$_{\text{H}_2}$ equilibria, we study the solubility of H$_2$ in an aqueous phase when it is in contact via a planar interface with a H$_2$ hydrate phase (H--L$_{\text{w}}$). We obtain the H$_2$ solubility values at $185\,\text{MPa}$ from 250 to $270\,\text{K}$ by running $NPT$ simulations. The barostat is applied anisotropically in the three space directions to avoid stress from the solid H$_2$ hydrate structure. Simulations are run for $800\,\text{ns}$, being the first $200\,\text{ns}$ considered as the equilibration period and the last $600\,\text{ns}$ as the production period.
In this case, the initial simulation box is prepared as follows. The hydrate part is formed from $1088$ molecules of water and a different number of H$_{2}$ molecules depending on the occupation level (see Table II). As can be seen, this corresponds to a $2\times2\times2$ unit cell of sII hydrate structure with different occupancies and space groups of the unit cell Fd3m. The proton disorder is obtained using the algorithm of Buch \emph{et al.}~\cite{Buch1998a} Once the hydrate phase is equilibrated, it is put in contact with a water liquid  phase.
The initial simulation box is built by a H$_2$ hydrate phase with 1088 molecules of water and the corresponding number of H$_2$ molecules at each occupancy studied in this work (see Table \ref{molecules}) in contact with a liquid aqueous phase. Notice that 1088 water molecules correspond to multiply twice the sII hydrate unit cell in each space direction (usually called as $2\times2\times2$ sII hydrate phase).
Due to the low solubility of H$_2$ in water, the initial aqueous phase in contact with the H$_2$ hydrate phase is built up as an initial pure water phase. Both phases are in contact via a planar interface along the $z$--direction. In order to reach the equilibrium solubility of H$_2$, part of the initial hydrate phase has to melt and release H$_2$ into the aqueous phase. This procedure is far from being arbitrary since it ensures that the hydrate phase is not going to grow along the simulation and, hence, the hydrate phase that remains in the simulation box, once the system has reached the equilibrium,  has the same occupancy (stoichiometry) as the initial one. This procedure has been recently used by some of the authors of this work to study the three-phase dissociation temperature of the N$_2$ hydrate when it presents single (1-1) and double (1-2) occupancies.~\cite{Torrejon2024b} 

From the analysis of the density profiles is possible to determine, not only the H$_2$ solubility in the aqueous phase but also whether the hydrate phase melts or grows. As has been remarked previously, the initial aqueous phase only contains water molecules and the only way to reach the equilibrium H$_2$ solubility is that part of the H$_2$ hydrate melts. Technically, the H$_2$ hydrate phase can not grow since there is no H$_2$ in the initial aqueous phase and an empty hydrate phase is not stable. However, when the occupancy of H$_2$ molecules in the hydrate is high enough and part of the hydrate melts, it could release into the aqueous phase more H$_2$ than necessary to reach the solubility of equilibrium, oversaturating the aqueous phase with H$_2$. This happens as a consequence of the low solubility of H$_2$ in the aqueous phase and the high amount of H$_2$ stored in the hydrate structure. When this happens, the hydrate phase grows to take back the excess of H$_2$ from the aqueous phase. If the hydrate phase grows, we can not ensure that the stoichiometry of the new hydrate phase is the same as the initial one and, hence, we can not ensure that the solubility of H$_2$ in the aqueous phase is reached when the aqueous phase is in contact with a H$_2$ hydrate phase at only the desire H$_2$ hydrate occupancy. To avoid the oversaturation of the aqueous phase when the H$_2$ hydrate melts, the number of molecules of water in the initial aqueous phase increases when the occupancy of the H$_2$ hydrate is increased (see Table \ref{molecules} for further details). In order to ensure that the stoichiometric of the hydrate along the simulation is the same as the initial seed, we carefully monitor the hydrate phase evolution by the analysis of the density profiles along the simulation. Density profiles provide an accurate description of the system distribution. The peaked shape region of the hydrate density phase contrasts with the flat density region of the fluid phase. By monitoring the length of the hydrate phase, it is possible to determine if the hydrate phase grows, melts, or both at different simulation times. In all cases, the hydrate phase only melts, ensuring that the remaining hydrate phase has the same stoichiometric as the initial seed. At this point is also important to mention that there is no diffusion of the guest through the hydrate phase and, as a consequence of this, the only option to modify the initial stoichiometric of the hydrate is the growing of this one with a different occupancy level. Also, it is important to take into account that the H$_2$ solubility in the aqueous phase in contact via a planar interface with the H$_2$ hydrate phase increases with the temperature. It means that when the temperature increases, and if the initial aqueous phase is big enough, the H$_2$ hydrate phase could melt completely in order to reach the equilibrium H$_2$ solubility value. For this reason, the number of water molecules in the initial aqueous phase is not the same at different temperatures even when the occupancy of the H$_2$ hydrate phase is the same (see, for example, the simulation boxes used for 2-2 and 1-4 occupancies in Table \ref{molecules}). 
Obviously, an increase in the number of molecules involves an increase in the computational effort required to perform the simulations. Hence, the number of water molecules in the aqueous phase is only increased if it is strictly necessary to ensure the stochiometry of the H$_2$ hydrate phase.

\begin{table}
\caption{\label{molecules}Initial number of molecules of water and H$_2$ in each H--L$_{\text{w}}$ simulation box at each H$_2$ hydrate occupancy and temperature. In all cases, the initial L$_{\text{w}}$ phase only contains water molecules.}

\begin{tabular}{c c c c c c c c c c c c c c c c c}

\hline
\hline
H$_2$ hydrate & & \multirow{2}{*}{$T$ (K)} & & \multicolumn{3}{c}{Hydrate phase} & &
\multicolumn{1}{c}{L$_{\text{w}}$ phase}  \\
     \cline{5-7} \cline{7-8} 
     occupancy & & & & Unit Cell  & Water & H$_2$ & & Water &     \\
      \hline
      \multirow{3}{*}{1-1} & & 250 & &\multirow{3}{*}{$2\times2\times2$}&\multirow{3}{*}{1088}&\multirow{3}{*}{192}& &\multirow{3}{*}{2176}&\\
       & &260 &&&&& &\\
       & &265 &&&&& &\\
      \hline
      \multirow{4}{*}{1-2} & &250& &\multirow{4}{*}{$2\times2\times2$}&\multirow{4}{*}{1088}&\multirow{4}{*}{256}& &\multirow{4}{*}{2176}&\\
       & &260& &&&& &\\
       & &265& &&&& &\\
       & &270& &&&& &\\
      \hline
      \multirow{4}{*}{2-2} & &250& &\multirow{4}{*}{$2\times2\times2$}&\multirow{4}{*}{1088}&\multirow{4}{*}{384}& &6528&\\
       & &260& &&&& &6528\\
       & &265& &&&& &6528\\
       & &270& &&&& &4352\\
      \hline
      \multirow{3}{*}{1-3} & &250& &\multirow{3}{*}{$2\times2\times2$}&\multirow{3}{*}{1088}&\multirow{3}{*}{320}& &4352&\\
       & &260& &&&& &4352\\
       & &265& &&&& &4352\\
      \hline
      \multirow{4}{*}{1-4} & &250& &\multirow{4}{*}{$2\times2\times2$}&\multirow{4}{*}{1088}&\multirow{4}{*}{384}& &6528&\\
       & &260& &&&& &6528\\
       & &265& &&&& &6528\\
       & &270& &&&& &4352\\
      \hline
      \hline
\end{tabular}
\end{table}

\begin{figure}
\includegraphics[width=\columnwidth]{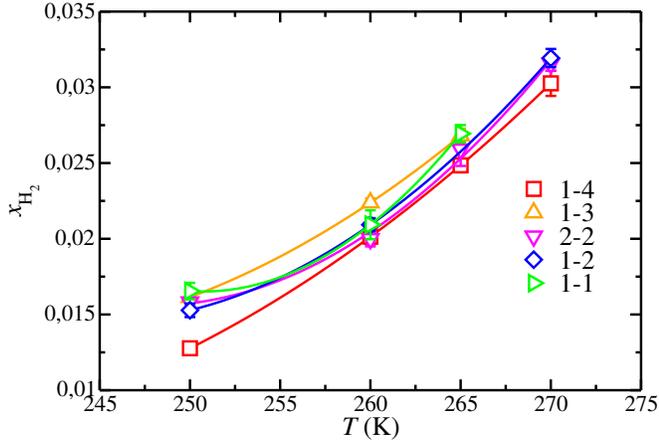}\\
\caption{H$_2$ solubility in the aqueous solution phase as a function of the temperature at $185\,\text{MPa}$. The symbols represent the solubility values obtained from the H--L$_{\text{w}}$ $NPT$ molecular dynamic simulations performed in this work at different H$_2$ hydrate occupancies. The meaning of the symbols is included in the legend. Each curve is obtained from the fitting of the solubility value at each occupancy and is included as a guide to the eyes.}
\label{figure3}
\end{figure}

The H$_2$ solubility in the aqueous phase is obtained following the same procedure as in the case of the L$_{\text{w}}$--L$_{\text{H}_2}$ equilibria. First, the density profiles of water and H$_2$ are obtained from each initial H$_2$ hydrate occupancy and temperature by dividing the simulation box into 200 slabs along the direction perpendicular to the interface (the $z$ axis). The center of mass of each molecule is assigned to its corresponding slab and the mass density profile is obtained by multiplying the molar density profile of each component by its corresponding molar mass. As it has been explained previously, the H$_2$ solubility in the aqueous phase at each H$_2$ hydrate occupancy and temperature is obtained by averaging the water and H$_2$ density profile values of each slab belonging to the aqueous phase. The error bars of the water and H$_2$ densities in the aqueous phase at each temperature are obtained as the average standard deviation. The H$_2$ solubility error bars are obtained by propagating the density errors obtained for both components. Again, calculations are performed far enough from the H--L$_{\text{w}}$ interface to avoid any unwanted interfacial effect on the aqueous bulk solubility of H$_2$. The results obtained in this work for each H$_2$ hydrate occupancy are presented in Fig. \ref{figure3}. As we can see, the solubility of H$_2$ in the aqueous phase when in contact with a H$_2$ hydrate phase via a planar interface increases when the temperature is increased. This is the expected behavior when the temperature increases since the hydrate phase becomes less stable and part of it melts releasing more H$_2$ into the aqueous phase. The solubility of H$_2$ in the aqueous phase, when it is in contact with a hydrate phase, can be understood as a special case of the solubility of a solid in water. Typically, the solubility of a solid in water increases when the temperature increases. In the case of hydrates, the same behavior is expected. However, hydrates are a special case since when part of their structure melts into the aqueous phase, two components are released, water and the guest. Since the solvent is water, the molecules of water from the hydrate become part of the proper solvent of the aqueous solution while the guest molecules become the solute. As we explained previously, the initial L$_{\text{w}}$ phase is basically a pure water phase, which means that all the H$_2$ present in this phase comes from the H$_2$ hydrate phase. It is interesting to remark that the solubility of H$_2$ in the aqueous phase seems to be almost independent of the H$_2$ hydrate occupancy since all the systems studied in this work present similar H$_2$ solubility values. It is also interesting to point out that, in the range of temperatures studied in this work, the H$_2$ solubility increases linearly when the temperature is increased.

\subsubsection{Three phase coexistence line ($T_3$) determination from solubility method}

\begin{figure}
\includegraphics[width=\columnwidth]{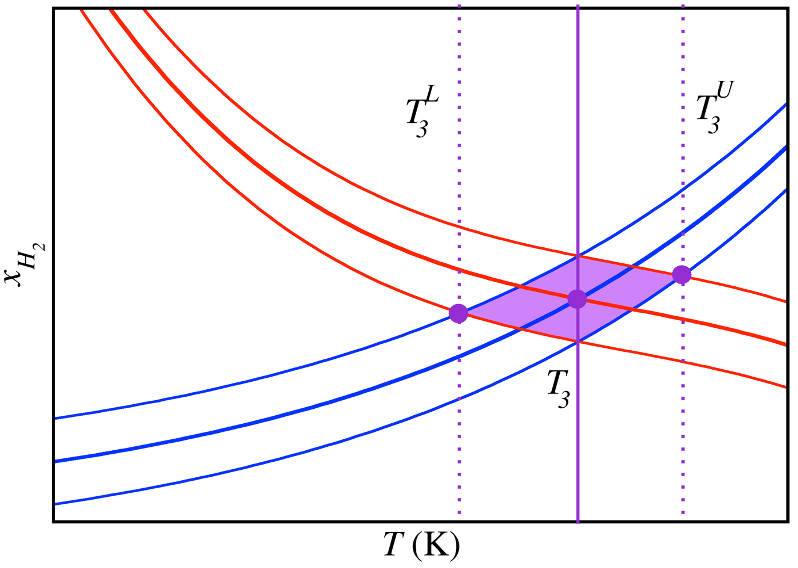}\\
\caption{Schematic representation of the determination of the $T_3$ value and its error bar. The three red and blue curves represent the solubility of H$_2$ in the aqueous phase as a function of the temperature from the L$_{\text{w}}$-L$_{\text{H}_2}$ and H--L$_{\text{w}}$ equilibria respectively. In both cases, the central red/blue line represents the equilibrium H$_2$ solubility, and the upper and lower red/blue lines the H$_2$ solubility error bars. The violet-filled circles represent the lowest ($T_3^L$), average ($T_3$), and upper ($T_3^U$) H$_2$ solubility and temperature at which the H$_2$ solubility curves cross.}
\label{figure4}
\end{figure}

\begin{figure}
\includegraphics[width=\columnwidth]{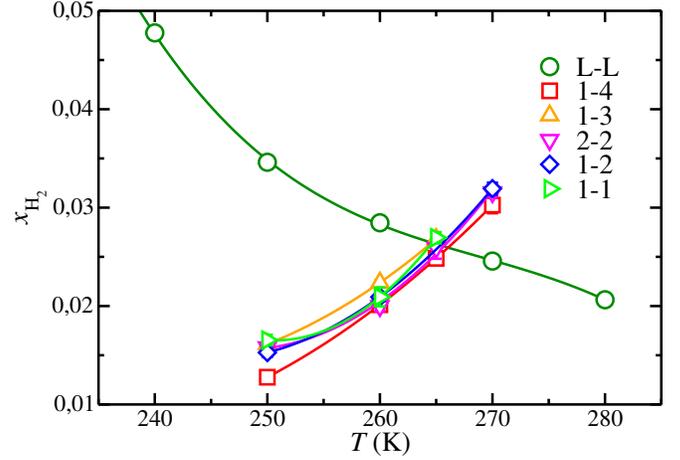}\\
\caption{$T_3$ values obtained at each H$_2$ hydrate occupancy as obtained from the solubility method. The meaning of the symbols is shown in the legend and it is the same as in the previous figures. The solubility curves are obtained by fitting the solubility H$_2$ values in the aqueous phase and are included as a guide to the eyes.}
\label{figure5}
\end{figure}

From the analysis of the solubility of H$_2$ in the aqueous phase obtained from the L$_{\text{w}}$-L$_{\text{H}_2}$ and H--L$_{\text{w}}$ equilibria, it is possible to determine the temperature ($T_3$) at which the three phases (H--L$_{\text{w}}$-L$_{\text{H}_2}$) coexist in equilibrium. The $T_3$ value is obtained as the temperature at which both H$_2$ solubility curves cross at a certain pressure value. It is also possible to get an estimation of the $T_3$ error bar by taking into account the error bars of the H$_2$ solubility values obtained from both (L$_{\text{w}}$-L$_{\text{H}_2}$ and H--L$_{\text{w}}$) equilibria. Figure \ref{figure4} shows a representation of the method employed in this work to determine the error bar of the $T_3$ values. For each equilibrium (L$_{\text{w}}$-L$_{\text{H}_2}$/H--L$_{\text{w}}$), three H$_2$ solubility curves are represented (red/blue). The central one represents the average H$_2$ solubility value obtained at each temperature, while the upper/lower ones are obtained by summing/resting to the average H$_2$ solubility value its corresponding error. As a result, we obtain a central $T_3$ value which corresponds to the temperature at which the central H$_2$ solubility curves from both equilibria intersect, a lower $T_3^L$ value obtained as the intersection of the upper and lower H$_2$ solubility curves obtained from the H--L$_{\text{w}}$ and L$_{\text{w}}$-L$_{\text{H}_2}$ equilibria respectively, and an upper $T_3^U$ value obtained as the intersection of the lower and upper H$_2$ solubility curves obtained from the H--L$_{\text{w}}$ and L$_{\text{w}}$-L$_{\text{H}_2}$ equilibria respectively. Finally, the $T_3$ error bar is obtained as $(T_3^U-T_3^L)/2$. The results obtained for each occupancy are collected in Table \ref{Table1} and the intersections of the solubility curves are presented in Fig. \ref{figure5}.

\begin{table}
\caption{Dissociation temperature, $T_3$, of the H$_{2}$ hydrate, at different H$_2$ hydrate occupancies, as obtained in this work using the solubility method at $185\,\text{MPa}$. Numbers in parentheses indicate the uncertainty of the results.}
\label{Table1}
\centering
\begin{tabular}{ccc}
\hline\hline
H$_2$ hydrate occupancy & & $T_3$ (K) \\
\hline
1-1 & & 264.6(6) \\
1-2 & & 265.4(7) \\
2-2 & & 265.6(8) \\
1-3 & & 264.6(7) \\
1-4 & & 266.1(8) \\
\hline\hline
\end{tabular}
\end{table}

As shown in Fig. \ref{figure5} and in Table \ref{Table1}, the H$_2$ hydrate occupancy has an almost negligible effect on the final $T_3$ value. All the $T_3$ values reported in this work at $185\,\text{MPa}$ are the same, within the error bars, independently of the H$_2$ hydrate occupancy. The $T_3$ values obtained for each occupancy at $185\,\text{MPa}$ are in agreement with that reported by Michalis \emph{et al.}~\cite{Michalis2022a} obtained using the TIP4P/Ice\cite{Abascal2005a} and Feynman-Hibbs\cite{Sese1993a} (FH) models for water and H$_2$ molecules respectively. Although the H$_2$ model (FH) employed in the original work of Michalis \emph{et al.}~\cite{Michalis2022a} to determine the H$_2$ hydrate $T_3$ value is different from the H$_2$ model used in this work (SG\cite{Silvera1978a}), the same modifying factor of the Berthelot rule (Eqs. (\ref{unlike}) and (\ref{chi})) for the water-H$_2$ interactions has been employed in both works. This is so since Michalis \emph{et al.}~\cite{Michalis2022a} demonstrated that both H$_2$ (FH and GS) models predict almost the same solubility of H$_2$ in an aqueous phase, modeled by the TIP4P/Ice water model, when in contact with a pure H$_2$ phase via a planar interface and the same modification of the Berthelot combining rule is required to match the experimental H$_2$ solubility results. As well as in the work of Michalis \emph{et al.}~\cite{Michalis2022a}, the $T_3$ values obtained in this study underestimate the experimental\cite{Dyadin1999a,Efimchenko2009a} $T_3$ values by $5-7\,\text{K}$. This is the expected result since it is similar to that obtained by Michalis \emph{et al.}~\cite{Michalis2022a} using the same modification of the Berthelot combining rule and the effect of the occupancy of the hydrate on the $T_3$ value has been demonstrated in this work to be almost negligible.

As we are going to analyze in the next section, although all the H$_2$ hydrate occupancies present the same $T_3$ value, not all of them seem to be equally favored thermodynamically. Technically, it is necessary to calculate the free energy of each hydrate at the three-phase coexistence conditions to analyze which occupancy presents the lowest free energy value and, hence, which one is the most stable. However, it is possible to estimate which occupancy is the most favored one by analyzing the driving force for nucleation behavior for each H$_2$ hydrate occupancy. Although the driving force for nucleation represents a difference of free energies and not an absolute free energy value, it gives us an approximated method to determine which H$_2$ hydrate occupancy is the most favored under the thermodynamic conditions at which this study has been carried out.

\subsection{Driving force for nucleation $\Delta\mu^{\text{EC}}_{N}$}

According to the Classical Nucleation Theory (CNT),~\cite{Debenedetti1996a} homogeneous nucleation is an activated process, which means that in order to crystallize the system has to overcome a free-energy barrier. This free-energy barrier depends on the hydrate-aqueous solution interfacial free energy, $\gamma$, and the driving force for nucleation, $\Delta\mu_{\text{N}}$. The driving force for nucleation is defined as the difference between the chemical potential of water and H$_2$ molecules in the hydrate phase, and those in the aqueous solution phase. At the thermodynamic equilibrium conditions of temperature, pressure, and composition, the value of $\Delta\mu_{\text{N}}$ is $0$ since the chemical potential of a molecule in each of the phases in equilibrium is the same. In the case of hydrates, the $\Delta\mu_{\text{N}}$ value becomes negative when the aqueous phase is supersaturated with the guest ~\cite{Walsh2011a,Liang2011a,Yagasaki2014a,Bagherzadeh2015a,Grabowska2022a,Fang2023a} (H$_2$ in this case) and when the temperature is below the dissociation temperature.~\cite{Kashchiev2002a,Kashchiev2002b,Debenedetti1996a} When $\Delta\mu_{\text{N}}$ becomes more negative, the free-energy barrier that the system has to overcome to crystallize becomes smaller, i.e. it is possible to favorer the homogeneous nucleation under supersaturation and/or supercooling conditions.

As we stated in our previous works,~\cite{Algaba2023a,Algaba2023b,Torrejon2024b} it is more convenient to express $\Delta\mu_{\text{N}}$ per cage of hydrate formed from the aqueous solution rather per guest molecule when the driving
forces for nucleation of hydrates with different occupancies are compared. In previous works, some of the authors of this work presented a general expression for semi/multiple occupied hydrates.~\cite{Algaba2023a,Algaba2023b,Torrejon2024b} Following the same approach, the occupancy of the H$_2$ hydrate is defined as $x_{\text{occ}}=n_{\text{H}_{2}}/n_{\text{cg}}$, where $n_{\text{H}_{2}}$ and $n_{\text{cg}}$ are the number of H$_{2}$ molecules and cages per hydrate unit cell, respectively.~\cite{Algaba2023a,Torrejon2024b} According to Kashchiev and Firoozabadi~\cite{Kashchiev2002a,Kashchiev2002b,Kashchiev2003a}, it is possible to describe the formation of a hydrate molecule from the aqueous solution phase as a chemical reaction at a certain value of $P$ and $T$. This approach could be generalized to any H$_2$ hydrate occupancy:

\begin{equation}
x_{\text{occ}}\,\text{H}_{2} (\text{aq},x_{\text{H}_{2}}) +
5.67\,\text{H}_{2}\text{O} (\text{aq},x_{\text{H}_{2}}) 
\rightarrow [(\text{H}_{2})_{x_{\text{occ}}}(\text{H}_{2}\text{O})_{5.67}]_{\text{H}}
\label{reaction}
\end{equation}

\noindent
$\text{H}_{2}(\text{aq},x_{\text{H}_{2}})$ and $\text{H}_{2}\text{O} (\text{aq},x_{\text{H}_{2}})$ are the molecules of H$_{2}$ and water in the aqueous solution phase with composition $x_{\text{H}_{2}}$, respectively, while the $[(\text{H}_{2})_{x_{\text{occ}}}(\text{H}_{2}\text{O})_{5.67}]_{\text{H}}$ represents a ``molecule'' of the H$_2$ hydrate in the solid phase with a certain  occupancy, $x_\text{occ}$. The $5.67$ factor arises because the sII hydrate unit cell is formed by $136$ molecules of water and $24$ cages and the "chemical reaction" of the H$_2$ hydrate formation is reduced according to the number of hydrate cages. The same approach is used to calculate the molar enthalpy of a "hydrate molecule", $\tilde{h}_{\text{H}}=H/N_{\text{cg}}$, where $H$ and $N_\text{cg}$ are the total enthalpy of the H$_2$ hydrate phase and the number of cages present in the hydrate structure, respectively. Although Eq. (\ref{reaction}) is a general expression and can be applied to any thermodynamic condition, it is particularly interesting to apply this expression along the L$_\text{w}$--L$_{\text{H}_2}$ solubility curve since most of the experiments on the nucleation of hydrates are performed when the water phase is in contact with the guest liquid phase through a planar interface. Following the notation of Grabowska and coworkers,~\cite{Grabowska2022a} the driving force for nucleation at experimental conditions is denoted as $\Delta\mu^{\text{EC}}_{\text{N}}$. Also, it is important to take into account that under experimental conditions, the $x_{\text{H}_{2}}$ value of Eq. (\ref{reaction}), at a given $P$ value, is a function of the temperature. In this work, $x_{\text{H}_{2}}$ has been obtained at $185\,\text{MPa}$ and different temperatures (see Fig. \ref{figure1}).

From Eq. (\ref{reaction}) it is possible to define an approximated but simple method to calculate $\Delta\mu^{\text{EC}}_{\text{N}}$.~\cite{Kashchiev2002a,Grabowska2022a,Algaba2023a,Algaba2023b,Torrejon2024b} If Eq. (\ref{reaction}) represents the formation of a H$_2$ hydrate molecule, the inverse process corresponds to the dissociation of a H$_2$ hydrate molecule into water and H$_2$ molecules. Due to the low solubility of H$_2$ in the aqueous phase, we can assume that the H$_2$ hydrate molecules dissociate into pure water and H$_2$. Taking this into account, the dissociation enthalpy of a hydrate molecule, $\tilde{h}_{\text{H}}^{\text{diss}}$, is defined as the enthalpy change of the hydrate dissociation in pure water and H$_2$ molecules. In order to calculate $\Delta\mu^{\text{EC}}_{\text{N}}$ in a simple way following the dissociation route approach,~\cite{Kashchiev2002a,Grabowska2022a,Algaba2023a,Algaba2023b, Torrejon2024b} extra approximations have to be taken into account: (1) the solubility of H$_2$ in the aqueous phase is assumed as 0 independently of the thermodynamic conditions, (2) $\tilde{h}_{\text{H}}^{\text{diss}}$ it is calculated at the $T_3$ value, and (3) it is considered as a constant value independently of the temperature. Taking all these approximations it is possible to define $\Delta\mu_{N}^{\text{EC}}$ as:

\begin{align}
\Delta\mu_{N}^{\text{EC}}(P,T,x_{\text{occ}})=k_{B}T
\bigintsss_{T_{3}}^{T}\dfrac{\tilde{h}^{\text{diss}}_{\text{H}}(P,T',x_{\text{occ}})}{k_{B}T'^{2}}\,dT' 
\nonumber\\ 
\approxeq -\tilde{h}^{\text{diss}}_{\text{H}} (P,T_{3},x_{\text{occ}}) \bigg(1-\dfrac{T}{T_{3}}\bigg)
\label{h_dissoc}
\end{align}

\noindent Where $k_B$ is the Boltzmann constant. Here it is important to remark that the occupancy of the hydrate is taken into account by the $\tilde{h}_{\text{H}}^{\text{diss}}$ value according to Eq. (\ref{reaction}). Equation (\ref{h_dissoc}) has been named as the dissociation route in our previous works.~\cite{Grabowska2022a, Algaba2023a,Algaba2023b,Torrejon2024b} 

When the approximations assumed in Eq. (\ref{h_dissoc}) are not undertaken, the expression for the $\Delta\mu_{N}^{\text{EC}}$ calculation becomes more complex since the solubility of H$_2$ in the aqueous phase has to be explicitly taken into account as well as the change of $\tilde{h}_{\text{H}}^{\text{diss}}$ with $T$ and $P$. As a result, we obtained the following expression named as route 1 expression in our previous works:~\cite{Grabowska2022a, Algaba2023a,Algaba2023b,Torrejon2024b}

\begin{widetext}
\begin{align}
\dfrac{\Delta\mu^{\text{EC}}_{\text{N}}(P,T,x^{\text{eq}}_{\text{H}_{2}},x_{occ})}
{k_{B}T}=&
-\bigintss_{T_{3}}^{T} 
\dfrac{\tilde{h}_{\text{H}}^{\text{H}}(P,T',x_{occ})- \Big\{x_{occ}h_{\text{H}_{2}}(P,T')+5.67\,
h_{\text{H}_{2}\text{O}}(P,T')\Big\}}{k_{B}T'^{2}}dT'
\nonumber\\
& - 5.67\Big[k_{B}T\ln\{x^{\text{eq}}_{\text{H}_{2}\text{O}}(P,T)\}-k_{B}T_{3}\ln\{x^{\text{eq}}_{\text{H}_{2}\text{O}}(P,T_{3})\}\Big]
\label{driving_force_route1}
\end{align}
\end{widetext}

\noindent Here $\tilde{h}_{\text{H}}^{\text{H}}(P,T',x_{occ})$ represents the enthalpy per cage of the hydrate at $P$ and $T'$ with occupancy $x_{\text{occ}}$ and $h_{\text{H}_2}$ and $h_{\text{H}_2\text{O}}$ represent the molar enthalpy of pure H$_2$ and water respectively. We refer the lector to our previous works for further details.

\begin{figure}
\includegraphics[width=\columnwidth]{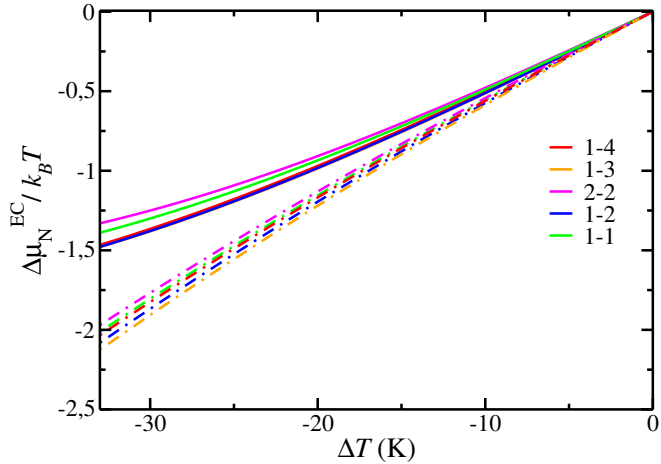}\\
\caption{$\Delta\mu^{\text{EC}}_{\text{N}}$ values for each H$_2$ hydrate occupancy studied in this work as a function of the supercooling degree at $185\,\text{MPa}$. Continuous and dashed lines represent the results obtained by Eqs. (\ref{driving_force_route1}) and (\ref{h_dissoc}) respectively.}
\label{figure6}
\end{figure}

Figure \ref{figure6} shows the $\Delta\mu^{\text{EC}}_{\text{N}}$ results obtained by both routes (Eqs. (\ref{driving_force_route1}) and (\ref{h_dissoc})) as a function of the supercooling degree for the different H$_2$ hydrate occupancies studied in this work. Although there are quantitative differences between the results obtained by both routes, the qualitative behavior of $\Delta\mu^{\text{EC}}_{\text{N}}$ with the supercooling degree and the occupancy is the same. For the same H$_2$ hydrate occupancy, $\Delta\mu^{\text{EC}}_{\text{N}}$ becomes more negative when the supercooling degree is increased. This is the expected result since the H$_2$ hydrate phase becomes more stable than the aqueous phase as the temperature decreases. Also, for the same supercooling degree, $\Delta\mu^{\text{EC}}_{\text{N}}$ becomes more negative when the amount of H$_2$ in the H, or large, cages increases from 1 to 3 (from 1-1 to 1-3 occupancy), being the 1-3 occupancy the most favored H$_2$ hydrate occupancy, closely followed by the 1-2 occupancy. However, if the number of H$_2$ molecules in the H cages is bigger than 3, then the H$_2$ hydrate phase becomes slightly less favored and, as a consequence, $\Delta\mu^{\text{EC}}_{\text{N}}$ increases becoming more positive. These results are in good agreement with those reported previously in the literature where the occupancy of the D and H cages of the H$_2$ hydrate was studied from Monte Carlo\cite{Brumby2019a,Katsumasa2007a,Chun2008a,Papadimitriou2008a} and molecular dynamic\cite{Alavi2005a} simulations. In Fig. \ref{figure6} we can see that the $\Delta\mu^{\text{EC}}_{\text{N}}$ results obtained, from both routes, for the 1-4 occupancy are between those obtained for the 1-1 and 1-2 occupancies. Also, we can state from Fig. \ref{figure6} that the double occupancy of the D, or small, cages destabilizes the hydrate structure due to the high repulsion present when two H$_2$ molecules are confined in the D cages. As a consequence, the $\Delta\mu^{\text{EC}}_{\text{N}}$ values obtained for the 2--2 occupancy are the most positive ones, revealing that among the different occupancies studied in this work, the 2--2 occupancy seems to be the less favored one. When considering hydrates as potential gas storage materials, the number of H$_2$ molecules per cage is a crucial factor in determining the storage capacity of a given hydrate structure. Experimentally, it is known that the sII hydrate small cages could host two H$_2$ molecules, while large cages could store up to four H$_2$ molecules,~\cite{Mao2002a}  leading to a total H$_2$ storage capacity of 5.0 wt.\% for H$_2$ hydrates. This H$_2$ storage capacity is close to the US DOE 5.5 wt.\% target for 2025.  However, our results indicate that the most stable hydrate configuration consists of three H$_2$ molecules in the large cages and only one H$_2$ molecule in the small one. This H$_2$ hydrate stoichiometric results in a storage capacity of 3.2 wt.\%, which is roughly bigger than half of the US DOE ultimate target.

\subsection{Effect of the temperature and occupancy on unit-cell size.}

\begin{figure}
\includegraphics[width=\columnwidth]{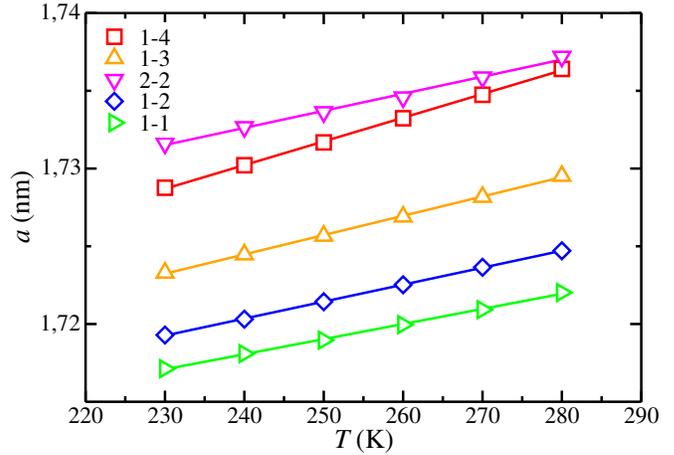}\\
\caption{Average unit-cell size or lattice constant ($a$) obtained from MD $NPT$ bulk simulations of the H$_2$ hydrate phase as a function of $T$ and the occupancy at $185\,\text{MPa}$. The meaning of the symbols is shown in the legend and it is the same as in previous figures. The lines are included as a guide to the eyes.}
\label{figure7}
\end{figure}

From the bulk simulations of the H$_2$ hydrate phase used to determine the $\Delta\mu^{\text{EC}}_{\text{N}}$, we analyze the average unit-cell size or lattice constant, $a$ as a function of the temperature and the occupancy of the H$_2$ hydrate phase at $185\,\text{MPa}$. Due to the cubic symmetry of the sII hydrate structure, we obtain $a$ as the cubic root of the average volume of the hydrate unit cell. As it is shown in Fig. \ref{figure7}, the H$_2$ hydrate lattice constant increases when the temperature is increased, although the dependency of $a$ with the temperature is almost negligible. This is an expected result since the same behavior was observed by some of the authors of this work for the case of the N$_2$ hydrate.~\cite{Torrejon2024b} It is also interesting to analyze the behavior of $a$ with the H$_2$ occupancy. As it is shown in Fig. \ref{figure7}, $a$ increases when the occupancy is increased. This is an expected behavior since a larger $a$ value allows to accommodate better the H$_2$ molecules inside the hydrate structure. As we can see, when the H$_2$ occupancy increases from 1 H$_2$ molecule in the H, or large, cages to 4 (i.e. from the 1-1 to 1-4 occupancies), the $a$ value is increased by $\approx0.7\%$ in all the range of temperatures. It is also interesting to observe that when the D, or small, and the H, or large, cages are doubly occupied (2-2 occupancy) the $a$ value is higher than in the 1-4 occupancy even when the total number of H$_2$ molecules in the H$_2$ hydrate unit cell is the same (48). This is so because the repulsion provoked by 2 H$_2$ molecules encapsulated in the small D and large H cages is larger than the repulsion provoked by 4 H$_2$ molecules in the large H cages and 1 H$_2$ molecule in the small D cages.

\subsection{Direct coexistence method}

Finally, we study the three-phase coexistence temperature, $T_3$, using the direct coexistence method. Following this methodology, the three phases involved in the equilibrium are placed together in the same simulation box. As has been explained previously, at a given pressure value, two different behaviors are shown as a function of the temperature. If the temperature is above the $T_3$ value, then the H$_2$ hydrate phase is unstable and it melts, evolving the three-phase system to a two-phase L$_\text{w}$--L$_{\text{H}_2}$ equilibrium. Contrary, if the temperature is below the $T_3$ value, then the aqueous phase becomes unstable and the H$_2$ hydrate phase grows until extinguishing the aqueous or guest phase depending on the initial amount of molecules of water and guest in both phases. To determine whether the H$_2$ hydrate phase grows or melts, the potential energy of the system is monitored. When the H$_2$ hydrate phase grows, new hydrogen bonds are created and the potential energy decreases. Contrary, when the H$_2$ hydrate phase melts the potential energy increases as a function of time. The $T_3$ is placed between the highest temperature at which the hydrate phase grows and the lowest at which it melts.

In this work, the initial configuration box is built up putting in contact the three phases, H$_{2}$ hydrate, aqueous, and pure H$_{2}$. The initial sII H$_{2}$ hydrate phase is built as explained in Section III.A.2. Particuarly, it is formed by a $2\times2\times2$ unit cell with an 1-3 H$_2$ occupancy for the hydrate phase in contact with an aqueous phase with 1088 water molecules and with a H$_2$ phase with 600 H$_2$ molecules. Due to the use of periodic boundary conditions, this arrangement ensures that the three phases are in contact, with one of the phases surrounded by the other two. We choose the 1-3 occupancy since it has been shown that with the most negative values of $\Delta\mu^{\text{EC}}_{\text{N}}$ as a function of the supercooling degree at $185\,\text{MPa}$. As we explained previously, it means that the 1-3 occupancy, from a thermodynamic point of view, is the most favored stoichiometry. Simulations were carried out at 100, 185, and $300\,\text{MPa}$ and at different temperatures. As it has been explained in Section II, we propose in this work an extra parameter, $b_0$, for the Berthelot modified combination rule (see Eq.(\ref{chi})) proposed by Michalis \emph{et al.}~\cite{Michalis2022a} in order to improve the predictions of the $T_3$ values in the $100-300\,\text{MPa}$ pressure range. Although the modified factor, $\chi$, proposed by Michalis \emph{et al.}~\cite{Michalis2022a} gives an excellent prediction of the H$_2$ solubility in water and improves the $T_3$ predictions, the $T_3$ values obtained still underestimate the experimental results reported in the literature\cite{Dyadin1999a,Efimchenko2009a} by $5-7\,\text{K}$.

\begin{figure}
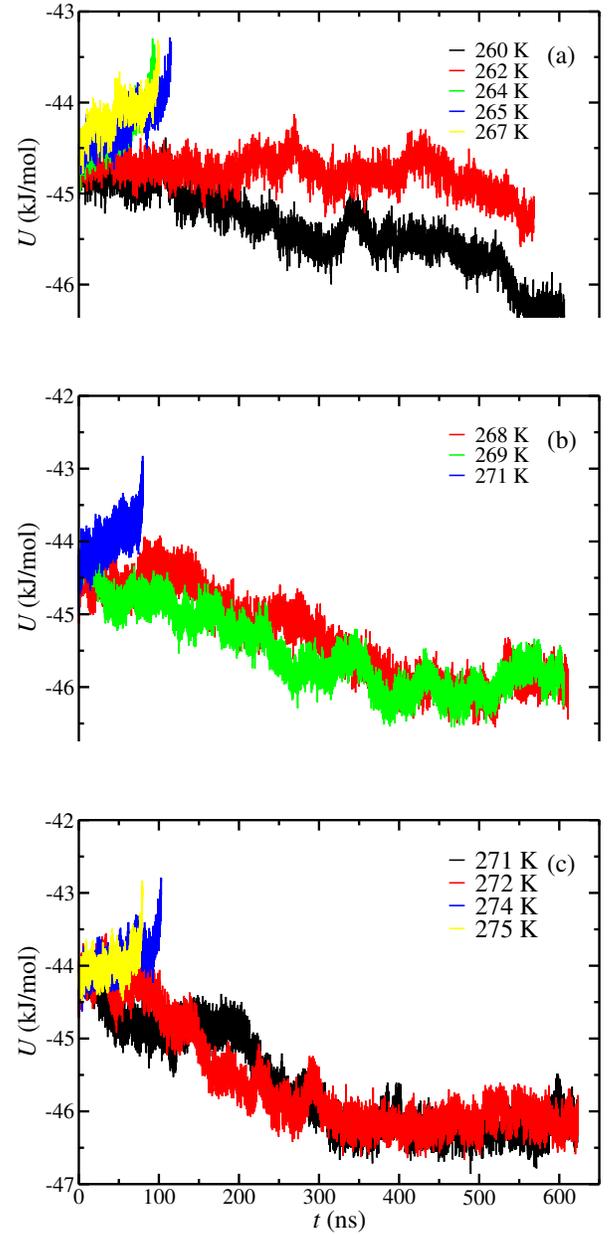

\centering
\includegraphics[width=0.9\columnwidth]{energies-1000bar.eps}
\hspace{0.2cm}
\includegraphics[width=0.9\columnwidth]{energies-1850bar.eps}
\hspace{0.2cm}
\includegraphics[width=0.90\columnwidth]{energies-3000bar.eps}
\caption{Potential energy as a function of time as obtained from $NPT$ molecular dynamic simulations at different temperatures and pressures. Figures (a), (b), and (c) correspond to results obtained at 100, 185, and $300\,\text{MPa}$ respectively. In all cases, the initial H$_2$ hydrate phase contains 3 H$_2$ molecules in the large H cages and 1 H$_2$ molecule in the small D cages (1-3 occupancy).}
\label{figure8}
\end{figure}

\begin{table}
\caption{Dissociation temperature of the H$_{2}$ hydrate obtained in this work from the direct coexistence technique ($T^{DC}_3$). Results from the experimental data reported in the literature\cite{Dyadin1999a,Efimchenko2009a} ($T^{EXP}_3)$ are also shown.}
\label{Table2}
\centering
\begin{tabular}{ccccc}
\hline\hline
$P$ (MPa) & & $T^{DC}_3$ (K) & &  $T^{EXP}_3$ (K)\\
\hline
100 & & 263(1) & &  264.1\\
185 & & 270(1) & & 269.9/271.6 \\
300 & & 273(1) & & 274.6\\
\hline\hline
\end{tabular}
\end{table}

\begin{figure}
\centering
\includegraphics[width=1.0\columnwidth]{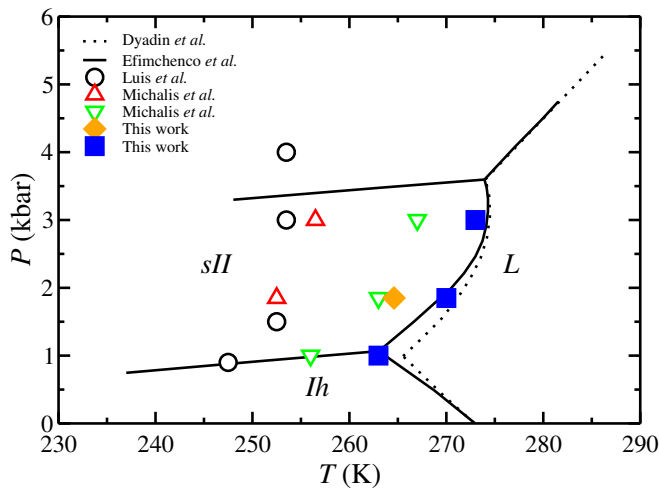}
\caption{H$_2$ Hydrate phase diagram. The sII, I$_h$, and L regions stand for sII H$_2$ hydrate, hexagonal ice, and liquid water respectively. Continuous and dotted black lines correspond to the experimental data reported by Efimchenco \emph{et al.}~\cite{Efimchenko2009a} and Dyadin \emph{et al.}~\cite{Dyadin1999a} respectively. Open symbols correspond to the simulation predictions of the three-phase coexistence line reported in the literature by Luis \emph{et al.}~\cite{Luis2018a} (black circles) and Michalis \emph{et al.}~\cite{Michalis2022a} (red up and green down triangles). Finally, filled symbols correspond to the three-phase coexistence conditions obtained in this work. The filled orange represents the predictions obtained by the solubility method at $185\,\text{MPa}$ and using the 1-3 occupancy and the original combining rules proposed by Michalis \emph{et al.}~\cite{Michalis2022a} (i.e., $b_0$=0 in Eq. \ref{chi}). Finally, the filled blue squares correspond to the predictions obtained using the direct coexistence technique and the extra parameter $b_0$=0.1.}
\label{figure9}
\end{figure}

Figure \ref{figure8} shows the potential energy results obtained at 100, 185, and $300\,\text{MPa}$ and different temperatures. From the results presented in Fig \ref{figure8} and following the procedure explained previously, it is easy to determine that 263(1), 270(1), and $273(1)\,\text{K}$ are the $T_3$ values obtained at 100, 185, and $300\,\text{MPa}$ respectively. As can be seen in Table \ref{Table2} and Fig. \ref{figure9}, the results obtained in this work using the direct coexistence technique and the additional $b_0$ parameter in Eq. (\ref{chi}) are equal, within the error bars, to the experimental data reported in the literature.~\cite{Dyadin1999a,Efimchenko2009a} It is also interesting to remark that the result obtained by the solubility method for the 1-3 H$_2$ hydrate occupancy is very close to the experimental data, slightly underestimating the experimental $T_3$ value at the same time that provides an accurate estimation of the H$_2$ solubility value in an aqueous phase. On the other hand, the addition of the $b_0$ parameter in Eq. (\ref{chi}) improves the prediction of the $T_3$ values but the H$_2$ solubility in an aqueous phase is slightly overestimated.~\cite{Michalis2022a} At this point is up to the lector to prioritize the prediction of the $T_3$ or the H$_2$ solubility by taking into account or not the $b_0$ parameter in Eq. (\ref{chi}). In both cases, the results obtained by both modified combining rules are excellent.

\section{Conclusions}

In this work, we have determined the three-phase H--L$_{\text{w}}$--L$_{\text{H}_{2}}$ coexistence line following two different molecular computer simulation methods. In both methods, the water and H$_2$ molecules are described using the TIP4P/Ice\cite{Abascal2005b} and a modified version of the SG model\cite{Alavi2005a,Silvera1978a,Michalis2022a} respectively. In both cases, the Berthelot combination rule for the water-H$_2$ interactions has been modified to improve the H$_2$ solubility and the $T_3$ predictions. Firstly, we study the effect of the H$_2$ hydrate occupancy on the $T_3$ value at $185\,\text{MPa}$ using the solubility method and 5 different H$_2$ hydrate occupancies (1-1, 1-2, 2-2, 1-3, and 1-4). We use the same modification of the Berthelot combining rule proposed by Michalis \emph{et al.}~\cite{Michalis2022a}, which provides an accurate estimation of the H$_2$ solubility in an aqueous phase but slightly underestimates the $T_3$ value. From this analysis, we conclude that the effect of the H$_2$ hydrate occupancy on the $T_3$ value is negligible since all the occupancies considered in this work provide almost the same $T_3$ result. The $T_3$ values obtained following this procedure are in very good agreement with the experimental data reported in the literature~\cite{Dyadin1999a,Efimchenko2009a} although they are slightly underestimated. 

We also estimate the driving force for nucleation, $\Delta\mu^{\text{EC}}_{N}$, of the H$_{2}$ hydrate as a function of the supercooling degree for the five different occupancies considered in this work. We find that multiple occupancy of the large H cages is favored in general, becoming $\Delta\mu^{\text{EC}}_{N}$ more negative when the occupancy is increased from 1 to 4 H$_2$ molecules, reaching the most negative value when the H cages are occupied by 3 H$_2$ molecules while the small D cages are single occupied. We also find that the double occupancy of the D and H cages is not favored, becoming $\Delta\mu^{\text{EC}}_{N}$ more positive than when both cages are singly occupied. This is so due to the high repulsion suffered by both H$_2$ molecules encapsulated inside the small D cages. As far as the authors know, this is the first time that the effect of the occupancy on the dissociation temperature and the driving force for nucleation of the H$_2$ hydrate is determined from computer simulations. 

Finally, we study the dissociation temperature at three different pressures (100, 185, and $300\,\text{MPa}$) using the direct coexistence technique. We perform the simulations with an initial 1-3 H$_2$ hydrate phase and, in order to improve the $T_3$ predictions, we propose a new modification of the Berthelot combining rule for the water-H$_2$ cross-interaction. The results obtained with this new combining rule modification are in excellent agreement with the experimental results reported in the literature.~\cite{Dyadin1999a,Efimchenko2009a}

\section*{Author declarations}

\noindent
\textbf{Conflict of interests}

The authors declare no conflicts to disclose.

\noindent

\section*{Data availability}

The data that support the findings of this study are available within the article.

\section*{Acknowledgements}

The authors wish to express their sincere gratitude to Dr. E. Dendy Sloan for his invaluable contributions to the study of gas hydrates. His pioneering work has not only been fundamental in advancing experimental knowledge of hydrates but has also opened new doors to molecular simulation in this field, enabling a deeper understanding of their structure and behavior. His legacy continues to guide both theoretical and applied research, inspiring future generations of scientists. We deeply appreciate his tireless dedication and the lasting impact of his contributions in both academia and industry. This work was financed by Ministerio de Ciencia e Innovaci\'on (Grants No.~PID2021-125081NB-I00 and PID2022-136919NB-C32) and Universidad de Huelva (P.O. FEDER EPIT1282023), both co-financed by EU FEDER funds. M. J. T. acknowledges the research contract (Ref.~01/2022/38143) of Programa Investigo (Plan de Recuperaci\'on, Transformaci\'on y Resiliencia, Fondos NextGeneration EU) from Junta de Andaluc\'{\i}a (HU/INV/0004/2022). We greatly acknowledge RES resources provided by Barcelona Supercomputing Center in Mare Nostrum to FI-2023-3-0011 and by The Supercomputing and Bioinnovation Center of the University of Malaga in Picasso to FI-2024-1-0017. S. B. acknowledges Ayuntamiento de Madrid for a Residencia de Estudiantes grant. The authors gratefully acknowledge the Universidad Politecnica de Madrid (www.upm.es) for providing computing resources on Magerit Supercomputer.

\section*{Bibliography}
\bibliography{masterbib}

@string{acs = {ACS Symposium Series.}}

@string{fuel = {Fuel}}

@string{jced = {J. Chem. Eng. Data}}

@string{jcg = {J. Cryst. Growth}}

@string{jcp = {J. Chem. Phys.}}

@string{jpc = {J. Phys. Chem.}}

@string{jpcb = {J. Phys. Chem. B}}

@string{jpcc = {J. Phys. Chem. C}}

@string{jsf = {J. Supercrit. Fluids}}

@string{molphys = {Mol. Phys.}}

@string{pnas = {Proc. Natl. Acad. Sci.}}

@string{pr = {Physics Reports}}

@string{science = {Science}}

@article{Miguez2015a,
Author = {J. M. M{\'\i}guez and M. M. Conde and J.-P. Torr{\'e} and F. J. Blas and M. M. Pi{\~n}eiro and C. Vega},
title={Molecular Dynamics Simulation of {CO$_2$} Hydrates: Prediction of Three Phase Coexistence Line},
Journal = jcp,
Pages = {124505},
Volume = {142},
doi={https://doi.org/10.1063/1.4916119},
Year = {2015}}

@article{deMiguel2006b,
	Author = {E. De Miguel and G. Jackson},
	Title={The Nature of the Calculation of the Pressure in Molecular Simulations of Continuous Models from Volume Perturbations},
	Journal = jcp,
	Pages = {164109},
	Volume = {125},
	Year = {2006}}

@article{deMiguel2006c,
Author = {E. De Miguel and G. Jackson},
title ={Detailed examination of the calculation of the pressure in simulations of systems with discontinuous interactions from the mechanical and thermodynamic perspectives},
journal = molphys,
Number = {22-24},
Pages = {3717--3734},
Volume = {104},
Year = {2006}}

@article{Essmann1995a,
Author = {U. Essmann and L. Perera and M. L. Berkowitz and T. Darden and H. Lee and L. G. Pedersen},
Title = {A Smooth Particle Mesh {Ewald} method},
Journal = jcp,
Pages = {8577-8593},
Volume = {103},
Year = {1995}}

@book{Rowlinson1982b,
	Author = {J. S. Rowlinson and B. Widom},
	Publisher = {Claredon Press},
	Title = {Molecular Theory of Capillarity},
	Year = {1982}}

@article{VanDerSpoel2005a,
Author = {D. {van der Spoel} and E. Lindahl and B. Hess and G. Groenhof and A. E. Mark and H. J. Berendsen},
Title={GROMACS: Fast, Flexible, and Free},
Journal = {J. Comput. Chem.},
Number = {16},
Pages = {1701-1718},
Volume = {26},
Year = {2005}}

@article{Nose1984a,
Author = {S. Nos{\'e}},
Title={A Molecular Dynamics Method for Simulations in the Canonical Ensemble},
Journal = molphys,
Pages = {255-268},
Volume = {52},
Year = {1984}}

@article{Cuendet2007a,
Author = {M. A. Cuendet and W. F. Van Gunsteren},
Title={On the Calculation of Velocity-Dependent Properties in Molecular Dynamics Simulations Using the Leapfrog Integration Algorithm},
Journal = jcp,
Pages = {184102},
Volume = {127},
Year = {2007}}

@book{Sloan2008a,
Address = {New York},
Author = {E. D. Sloan and C. Koh},
Edition = {3},
Publisher = {CRC Press},
Title = {{C}lathrate {H}ydrates of {N}atural {G}ases},
Year = {2008}}

@article{Torre2015a,
  title={1, 3 Dioxolane versus tetrahydrofuran as promoters for CO2-hydrate formation: Thermodynamics properties, and kinetics in presence of sodium dodecyl sulfate},
  author={Torr{\'e}, J-P and Haillot, Didier and Rigal, Sacha and de Souza Lima, Roger and Dicharry, Christophe and Bedecarrats, J-P},
  journal={Chemical Engineering Science},
  volume={126},
  pages={688--697},
  year={2015},
  publisher={Elsevier}}

@article{Mao2002a,
Author = {W. L. Mao and H. K. Mao and A. F. Goncharov and V. V. Struzhkin and Q. Guo
	and J. Hu and J. Shu and R. J. Hemley and M. Somayazulu and Y. Zhao},
Title={Hydrogen Clusters in Clathrate Hydrate},
Journal = science,
Pages = {2247-2249},
Volume = {297},
Year = {2002}}

@article{Mao2004a,
Author = {W. L. Mao and H. K. Mao},
Title={Hydrogen Storage in Molecular Compounds},
Journal = pnas,
Pages = {708-710},
Volume = {101},
Year = {2004}}

@article{Makino2005a,
Author = {T. Makino and T. Sugahara and K. Ohgaki},
Title={Stability Boundaries of Tetrahydrofuran + Water System},
Journal = jced,
Pages = {2058-2060},
Volume = {50},
Year = {2005}}

@article{Parrinello1981a,
Author = {M. Parrinello and A. Rahman},
Title={Polymorphic Transitions in Single Crystals: A New Molecular Dynamics Method},
Journal = {J. Appl. Phys.},
Pages = {7182-7190},
Volume = {52},
Year = {1981}}

@article{Conde2010a,
Author = {M. M. Conde and C. Vega},
  title={Determining the three-phase coexistence line in methane hydrates using computer simulations},
  journal=jcp,
  volume={133},
  number={6},
  pages={064507},
  year={2010},
  doi={https://doi.org/10.1063/1.3466751},
  publisher={American Institute of Physics}
}

@article{Conde2013a,
Author = {M. M. Conde and C. Vega},
  title={Note: A simple correlation to locate the three phase coexistence line in methane-hydrate simulations},
Journal = jcp,
Pages = {056101},
Volume = {138},
  number={5},
  publisher={American Institute of Physics},
  doi={https://doi.org/10.1063/1.4790647},
Year = {2013}}

@article{Abascal2005a,
Author = {J. L. Abascal and C. Vega},
title={A General Purpose Model for the Condensed Phases of Water: {TIP4P/2005}},
Journal = jcp,
Pages = {234505},
Volume = {123},
Year = {2005}}

@book{Debenedetti1996a,
Author = {P. G. Debenedetti},
Publisher = {Princeton University Press},
Title = {Metastable Liquids: Concepts and Principles},
Year = {1996}}

@article{Abascal2005b,
Author = {J. L. F. Abascal and E. Sanz and R. Garc{\'{\i}}a Fern\'andez and C. Vega},
Title={A Potential Model for the Study of Ices and Amorphous Water: {TIP4P/Ice}},
Journal = jcp,
Pages = {234511},
Volume = {122},
Year = {2005}}

@article{Grabowska2022a,
Author = {J. Grabowska and S. Bl{\'a}zquez and E. Sanz and I. M. Zer{\'o}n and
J. Algaba and J. M. M{\'{\i}}guez and F. J. Blas and C. Vega},
Title={Solubility of methane in water: some useful results for hydrate nucleation},
Journal = jpcb, 
Pages = {8553-8570},
Volume = {126},
Year = {2022}}

@article{Kashchiev2002a,
Author = {D. Kashchiev and A. Firoozabadi},
Title={Driving force for crystallization of gas hydrates},
Journal = jcg, 
Pages = {220-230},
Volume = {241},
Year = {2002}}

@article{Kashchiev2002b,
Author = {D. Kashchiev and A. Firoozabadi},
Title={Nucleation of gas hydrates},
Journal = jcg, 
Pages = {476-489},
Volume = {243},
Year = {2002}}

@article{Kashchiev2003a,
Author = {D. Kashchiev and G. M. van Rosmalen},
Title={Review: Nucleation in solutions revisited},
Journal = "Cryst. Res. Technol.", 
Pages = {555-574},
Volume = {38},
doi = {https://doi.org/10.1002/crat.200310070},
Year = {2003}}

@article{Michalis2015a,
  title={Prediction of the phase equilibria of methane hydrates using the direct phase coexistence methodology},
  author={V. K. Michalis and J. Costandy and I. N. Tsimpanogiannis and A. K. Stubos and I. G. Economou},
  journal={J. Chem. Phys.},
  volume={142},
  number={4},
  pages={044501},
  year={2015},
  doi={https://doi.org/10.1063/1.4905572},
  publisher={AIP Publishing LLC}
}

@article{Costandy2015a,
  title={The role of intermolecular interactions in the prediction of the phase equilibria of carbon dioxide hydrates},
  author={J. Costandy and V. K. Michalis and I. N. Tsimpanogiannis and A. K. Stubos and I. G. Economou},
  journal={J. Chem. Phys.},
  volume={143},
  number={9},
  pages={094506},
  year={2015},
  doi={https://doi.org/10.1063/1.4929805},
  publisher={AIP Publishing LLC}
}

@article{Perez-Rodriguez2017a,
  title={Computational study of the interplay between intermolecular interactions and {CO$_{2}$} orientations in type {I} hydrates},
  author={M. P{\'e}rez-Rodr{\'\i}guez and A. Vidal-Vidal and J.M. M{\'\i}guez and F. J. Blas and J-P. Torr{\'e} and M. M. Pi{\~n}eiro},
  journal={Phys. Chem. Chem. Phys.},
  volume={19},
  number={4},
  pages={3384--3393},
  year={2017},
  doi={https://doi.org/10.1039/C6CP07097C},
  publisher={Royal Society of Chemistry}
}

@article{Fernandez-Fernandez2019a,
  title={Three-phase equilibrium curve shift for methane hydrate in oceanic conditions calculated from Molecular Dynamics simulations},
  author={A. M. Fern{\'a}ndez-Fern{\'a}ndez and M. P{\'e}rez-Rodr{\'\i}guez and A. Comesa{\~n}a and M. M. Pi{\~n}eiro},
  journal={J. Mol. Liq.},
  volume={274},
  pages={426--433},
  year={2019},
  doi={https://doi.org/10.1016/j.molliq.2018.10.146},
  publisher={Elsevier}
}

@article{Walsh2011a,
Author = {M. R. Walsh and  G. T. Beckham and C. A. Koh and E. D. Sloan and D. T. Wu and A. K. Sum},
title={Methane Hydrate Nucleation Rates from Molecular Dynamics Simulations: Effects of Aqueous Methane Concentration, Interfacial Curvature, and System Size},
Journal = jpcc,
Pages = {21241},
Volume = {115},
  doi={https://doi.org/10.1021/jp206483q},
Year = {2011}}

@article{Algaba2023a,
  title={Solubility of carbon dioxide in water: Some useful results for hydrate nucleation},
  author={Algaba, Jes{\'u}s and Zer{\'o}n, Iv{\'a}n M and M{\'\i}guez, Jos{\'e} Manuel and Grabowska, Joanna and Blazquez, Samuel and Sanz, Eduardo and Vega, Carlos and Blas, Felipe J},
  journal={J. Chem. Phys.},
  volume={158},
  number={18},
  pages={054505},
  year={2023},
  publisher={AIP Publishing}
}

@article{Yi2019a,
  title={Molecular Dynamics Simulation Study on the Growth of Structure {II} Nitrogen Hydrate},
  author={Yi, Lizhi and Zhou, Xuebing and He, Yunbing and Cai, Zhuodi and Zhao, Lili and Zhang, Wenkai and Shao, Youyuan},
  journal={J. Phys. Chem. B},
  volume={123},
  number={43},
  pages={9180--9186},
  year={2019},
  publisher={ACS Publications}
}

@article{Brumby2019a,
  title={Cage occupancies, lattice constants, and guest chemical potentials for structure {II} hydrogen clathrate hydrate from {G}ibbs ensemble {M}onte {C}arlo simulations},
  author={P. E. Brumby and D. Yuhara and T. Hasegawa and D. T. Wu and A. K. Sum and K. Yasuoka},
  journal=jcp,
  volume={150},
  pages={134503},
  year={2019},
}

@article{Ladd1977a,
  title={Triple-point coexistence properties of the Lennard-Jones system},
  author={J. C. Ladd and L. V. Woodcock},
  journal={Chem. Phys. Lett.},
  volume={51},
  pages={159155},
  year={1977},
}

@article{Blazquez2023b,
  title={Three phase equilibria of the methane hydrate in {NaCl} solutions: A simulation study},
  author={S. Blazquez and C. Vega and M. M. Conde},
  journal={J. Mol. Liq.},
  volume={383},
  pages={122031},
  year={2023},
}

@article{Flyvbjerg1989a,
  title={Error estimates on averages of correlated data},
  author={H. Flyvbjerg and H.G. Petersen},
  journal=jcp,
  volume={91},
  pages={461-466},
  year={1989},
}

@article{Buch1998a,
  title={Simulations of {H$_{2}$O} Solid, Liquid, and Clusters, with an Emphasis on Ferroelectric Ordering Transition in Hexagonal Ice},
  author={V. Buch and P. Sandler and J. Sadlej},
  journal=jpcb,
  volume={102},
  pages={8641-8653},
  year={1998},
}

@article{Michalis2022a,
Author = {V. K. Michalis and I. G. Economou and A. K. Stubos and I. N. Tsimpanogiannis},
Title={Phase equilibria molecular simulations of hydrogen hydrates via the direct phase coexistence approach},
Journal = jcp,
Pages = {154501},
Volume = {157},
Year = {2022} 
}

@article{Tsimpanogiannis2017a,
Author = {I. N. Tsimpanogiannis and I. G. Economou},
Title={Monte Carlo simulation studies of clathrate hydrates: A review},
Journal = jsf,
Pages = {51-60},
Volume = {134},
Year = {2017}
}

@article{Alavi2005a,
Author = {S. Alavi and J. A. Ripmeester and D. D. Klug},
Title={Molecular-dynamics study of structure {II} hydrogen clathrates},
Journal = jcp,
Pages = {024507},
Volume = {123},
Year = {2005}
}

@book{Ripmeester2022a,
Author = {J. A. Ripmeester and S. Alavi},
Publisher = {Wiley-VCH: Weinheim, Germany},
Title = {Clathrate Hydrates: Molecular Science and Characterization},
Year = {2022}
}

@article{Algaba2023b,
  title={Dissociation line and driving force for nucleation of the nitrogen hydrate from computer simulation},
  author={Algaba, Jes{\'u}s and Torrej{\'o}n, Miguel J and Blas, Felipe J},
  journal={J. Chem. Phys.},
  volume={159},
  number={22},
  pages={224707},
  year={2023},
  publisher={AIP Publishing}
}

@article{dashti2015recent,
  title={Recent advances in gas hydrate-based {CO$_{2}$} capture},
  author={Dashti, Hossein and Yew, Leonel Zhehao and Lou, Xia},
  journal={J. Natural Gas Sci. Eng.},
  volume={23},
  pages={195--207},
  year={2015},
}

@article{cannone2021review,
  title={A review on {CO$_{2}$} capture technologies with focus on {CO$_{2}$}-enhanced methane recovery from hydrates},
  author={Cannone, Salvatore F and Lanzini, Andrea and Santarelli, Massimo},
  journal={Energies},
  volume={14},
  number={2},
  pages={387},
  year={2021},
}

@article{duc2007co2,
  title={{CO$_{2}$} capture by hydrate crystallization--{A} potential solution for gas emission of steelmaking industry},
  author={Duc, Nguyen Hong and Chauvy, Fabien and Herri, Jean-Michel},
  journal={Energy Convers. Manag.},
  volume={48},
  number={4},
  pages={1313--1322},
  year={2007},
}

@article{choi2022effective,
  title={Effective {CH$_{4}$} production and novel {CO$_{2}$} storage through depressurization-assisted replacement in natural gas hydrate-bearing sediment},
  author={Choi, Wonjung and Mok, Junghoon and Lee, Jonghyuk and Lee, Yohan and Lee, Jaehyoung and Sum, Amadeu K and Seo, Yongwon},
  journal={Appl. Energy},
  volume={326},
  pages={119971},
  year={2022},
  publisher={Elsevier}
}

@article{lee2014quantitative,
  title={Quantitative measurement and mechanisms for {CH$_{4}$} production from hydrates with the injection of liquid {CO$_{2}$}},
  author={Lee, Bo Ram and Koh, Carolyn A and Sum, Amadeu K},
  journal={Phys. Chem. Chem. Phys.},
  volume={16},
  number={28},
  pages={14922--14927},
  year={2014},
  publisher={Royal Society of Chemistry}
}

@article{ma2016review,
  title={Review of fundamental properties of {CO$_{2}$} hydrates and {CO$_{2}$} capture and separation using hydration method},
  author={Ma, ZW and Zhang, P and Bao, HS and Deng, S},
  journal={Renew. Sustain. Energy Rev.},
  volume={53},
  pages={1273--1302},
  year={2016},
}

@article{hassanpouryouzband2018co2,
  title={{CO$_{2}$} capture by injection of flue gas or {CO$_{2}$}--{N$_{2}$} mixtures into hydrate reservoirs: Dependence of {CO$_{2}$} capture efficiency on gas hydrate reservoir conditions},
  author={Hassanpouryouzband, Aliakbar and Yang, Jinhai and Tohidi, Bahman and Chuvilin, Evgeny and Istomin, Vladimir and Bukhanov, Boris and Cheremisin, Alexey},
  journal={Environ. Sci. Technol.},
  volume={52},
  number={7},
  pages={4324--4330},
  year={2018},
  publisher={ACS Publications}
}

@article{Asadi2019a,
  title={Investigation of THF hydrate formation kinetics: Experimental measurements of volume changes},
  author={Asadi, Masomeh and Peyvandi, Kiana and Varaminian, Farshad and Mokarian, Zahra},
  journal={Journal of Molecular Liquids},
  volume={290},
  pages={111200},
  year={2019},
  publisher={Elsevier}
}

@article{Sun2017a,
  title={Stochastic nature of nucleation and growth kinetics of THF hydrate},
  author={Sun, Shicai and Peng, Xia and Zhang, Yong and Zhao, Jie and Kong, Yayun},
  journal={The Journal of Chemical Thermodynamics},
  volume={107},
  pages={141--152},
  year={2017},
  publisher={Elsevier}
}

@article{Suzuki2011a,
  title={Foreign particle behavior at the growth interface of tetrahydrofuran clathrate hydrates},
  author={Suzuki, Takahiro and Muraoka, Michihiro and Nagashima, Kazushige},
  journal={Journal of crystal growth},
  volume={318},
  number={1},
  pages={131--134},
  year={2011},
  publisher={Elsevier}
}

@article{Sabase2009a,
  title={Growth mode transition of tetrahydrofuran clathrate hydrates in the guest/host concentration boundary layer},
  author={Sabase, Yuichiro and Nagashima, Kazushige},
  journal={J. Phys. Chem. B},
  volume={113},
  number={46},
  pages={15304--15311},
  year={2009},
  publisher={ACS Publications}
}

@article{Strauch2018a,
  title={The difference between aspired and acquired hydrate volumes--A laboratory study of THF hydrate formation in dependence on initial THF: {H$_{2}$O} ratios},
  author={Strauch, Bettina and Schicks, Judith M and Luzi-Helbing, Manja and Naumann, Rudolf and Herbst, Marcel},
  journal={The Journal of Chemical Thermodynamics},
  volume={117},
  pages={193--204},
  year={2018},
  publisher={Elsevier}
}

@article{Ganji2006a,
  title={A kinetic study on tetrahydrofuran hydrate crystallization},
  author={Ganji, Hamid and Manteghian, Mehrdad and Zadeh, Kambiz Sadaghiani},
  journal={Journal of chemical engineering of Japan},
  volume={39},
  number={4},
  pages={401--408},
  year={2006},
  publisher={The Society of Chemical Engineers, Japan}
}

@article{Andersson1996a,
  title={Thermal conductivity of normal and deuterated tetrahydrofuran clathrate hydrates},
  author={Andersson, O and Suga, H},
  journal={Journal of Physics and Chemistry of Solids},
  volume={57},
  number={1},
  pages={125--132},
  year={1996},
  publisher={Elsevier}
}

@article{Chong2016a,
  title={Review of natural gas hydrates as an energy resource: Prospects and challenges},
  author={Chong, Zheng Rong and Yang, She Hern Bryan and Babu, Ponnivalavan and Linga, Praveen and Li, Xiao-Sen},
  journal={Appl. Energy},
  volume={162},
  pages={1633--1652},
  year={2016},
  publisher={Elsevier}
}

@article{Kumar2010a,
  title={Experimental determination of permeability in the presence of hydrates and its effect on the dissociation characteristics of gas hydrates in porous media},
  author={Kumar, Anjani and Maini, Brij and Bishnoi, PR and Clarke, Matthew and Zatsepina, Olga and Srinivasan, Sanjay},
  journal={Journal of petroleum science and engineering},
  volume={70},
  number={1-2},
  pages={114--122},
  year={2010},
  publisher={Elsevier}
}

@article{Blazquez2024a,
  title={Three-phase equilibria of hydrates from computer simulation. {I}: {F}inite-size effects in the methane hydrate},
  author={S. Blazquez and J. Algaba and  J. M. Míguez and C. Vega and F. J. Blas and M. M. Conde},
  journal={J. Chem. Phys.},
  volume={160},
  pages={164721},
  number={16},
  year={2024},
 }

@article{Algaba2024a,
  title={Three-phase equilibria of hydrates from computer simulation. {II}: {F}inite-size effects in the carbon dioxide hydrate},
  author= {J. Algaba and S. Blazquez and E. Feria and J. M. Míguez and M. M. Conde and F. J. Blas},
  journal={J. Chem. Phys.},
  volume={160},
  number={16},
  pages={164722},
year={2024},
 }

@article{Algaba2024b,
  title={Three-phase equilibria of hydrates from computer simulation. {III}: {E}ffect of dispersive interactions in methane and carbon dioxide hydrates},
  author={J. Algaba and S. Blazquez and J. M. Míguez and M. M. Conde and F. J. Blas},
  journal={J. Chem. Phys.},
  volume={160},
  number={16},
  pages={164723},
  year={2024},
 }

@article{Bagherzadeh2015a,
  title={Formation of methane nano-bubbles during hydrate decomposition and their effect on hydrate growth},
  author={S. A. Bagherzadeh  and S. Alavi and J. Ripmeester and P. Englezos},
  journal={J. Chem. Phys.},
  volume={142},
  pages={214701},
  year={2015},
  publisher={AIP Publishing}
}

@Article{Yagasaki2014a,
author = "T. Yagasaki and M. Matsumoto and Y. Andoh and S. Okazaki and H. Tanaka",
title = " Effect of Bubble Formation on the Dissociation of Methane Hydrate in Water: A Molecular Dynamics Study",
journal = {J. Phys. Chem. B},
volume = "118 ",
pages = "1900",
year = " 2014",
}

@article{Fang2023a,
  title={Effects of nanobubbles on methane hydrate dissociation: A molecular simulation study},
  author={B. Fang and O. Moultos and T. L{\"u}and J. Sun and Z. Liu and F. Ning and T. J. H. Vlugt},
  journal={Fuel},
  volume={345},
  pages={128230},
  year={2023},
  publisher={Elsevier}
}

@article{Liang2011a,
title={Exploring nucleation of {H$_{2}$S} hydrates},
author={S. Liang and P. G. Kusalik},
journal={Chem. Sci.},
volume={2},
pages={1286--1292},
year={2011},
publisher={Royal Society of Chemistry}
}

@article{Vega2008a,
title={Determination of phase diagrams via computer simulation: Methodology and applications to water, electrolytes and proteins},
author={C. Vega and E. Sanz and J. L. F. Abacal and E. G. Noya},
journal={J. Phys.: Condensed Matter},
volume={20},
pages={153101},
year={2008},
}

@article{Algaba2024c,
title={Prediction of the univariant two-phase coexistence line of the tetrahydrofuran hydrate from computer simulation},
author={J. Algaba and C. Romero-Guzm{\'a}n and M. J. Torrej{\'o}n and F. J. Blas},
journal=jpc,
volume={160},
pages={164718},
year={2024},
}

@article{Zhang2023a,
author = {Z. Zhang and P. G. Kusalik and C. Liu and N. Wu},
title = {Methane hydrate formation in slit-shaped pores: Impacts of surface hydrophilicity},
journal = {Energy},
volume = {285},
pages = {129414},
year = {2023},
doi = {https://doi.org/10.1016/j.energy.2023.129414}
}

@article{Zhang2022a,
author = {Z. Zhang and P. G. Kusalik and N. Wu and C. Liu and Y. Zhang},
title = {Molecular simulation study on the stability of methane hydrate confined in slit-shaped pores},
journal = {Energy},
volume = {257},
pages = {124738},
year = {2022},
doi = {https://doi.org/10.1016/j.energy.2022.124738},
}

@article{Fernandez-Fernandez2024a,
title={Modeling oceanic sedimentary methane hydrate growth through molecular dynamics simulation},
author={A. M. Fern{\'a}ndez-Fern{\'a}ndez and A. B{\'a}rcena and  M. M. Conde and G. P{\'e}rez-S{\'a}nchez and M. P{\'e}rez-Rodr{\'{\i}}guez and M. M. Pi{\~n}eiro},
journal = jcp,
volume={160},
pages ={144107},
year = {2024},
doi={https://doi.org/10.1063/5.0203116}
}

@article{Zheng2020a,
  title = {Carbon dioxide sequestration via gas hydrates: a potential pathway toward decarbonization},
  author = {Zheng, Junjie and Chong, Zheng Rong and Qureshi, M Fahed and Linga, Praveen},
  journal = {Energy \& Fuels},
  volume = {34},
  number = {9},
  pages = {10529--10546},
  year = {2020},
  publisher = {ACS Publications}
}

@article{Bourry2007a,
  title={X-ray synchrotron diffraction study of natural gas hydrates from African margin},
  author={Bourry, Christophe and Charlou, Jean-Luc and Donval, Jean-Pierre and Brunelli, Michela and Focsa, Cristian and Chazallon, Bertrand},
  journal={Geophysical Research Letters},
  volume={34},
  number={22},
  year={2007},
  publisher={Wiley Online Library}
}

@article{Silvera1978a,
  title={The isotropic intermolecular potential for {H$_{2}$} and {D$_{2}$} in the solid and gas phases},
  author={Silvera, Isaac F and Goldman, Victor V},
  journal={The Journal of Chemical Physics},
  volume={69},
  number={9},
  pages={4209--4213},
  year={1978},
  publisher={American Institute of Physics}
}

@article{Torrejon2024a,
  title={Simulation of the {THF} hydrate - water interfacial free energy from computer simulation},
  author={M. J. Torrejón and C. Romero-Guzmán and M. M. Piñeiro and F. J. Blas and J. Algaba},
  journal=jcp,
  volume={161},
  pages={064701},
  year={2024}
}

@article{Luis2018a,
  title={The coexistence temperature of hydrogen clathrates: A molecular dynamics study},
  author={Luis, DP and Romero-Ramirez, IE and Gonz{\'a}lez-Calder{\'o}n, A and L{\'o}pez-Lemus, J},
  journal={The Journal of Chemical Physics},
  volume={148},
  number={11},
  year={2018},
  publisher={AIP Publishing}
}

@article{Torrejon2024b,
  title={Dissociation line and driving force for nucleation of the nitrogen hydrate from computer simulation. II. Effect of multiple occupancy},
  author={Torrej{\'o}n, Miguel J and Algaba, Jes{\'u}s and Blas, Felipe J},
  journal={The Journal of Chemical Physics},
  volume={161},
  number={5},
  year={2024},
  publisher={AIP Publishing}
}

@article{Sese1993a,
  title={Feynman-Hibbs quantum effective potentials for Monte Carlo simulations of liquid neon},
  author={Ses{\'e}, Luis M},
  journal={Molecular Physics},
  volume={78},
  number={5},
  pages={1167--1177},
  year={1993},
  publisher={Taylor \& Francis}
}

@article{Dyadin1999a,
  title={Clathrate hydrates of hydrogen and neon},
  author={Dyadin, Yuri A and Larionov, Eduard G and Manakov, Andrei Y and Zhurko, Fridrich V and Aladko, Eugeny Y and Mikina, Tamara V and Komarov, Vladislav Y},
  journal={Mendeleev communications},
  volume={9},
  number={5},
  pages={209--210},
  year={1999},
  publisher={Royal Society of Chemistry}
}

@article{Efimchenko2009a,
  title={Two triple points in the H2O--H2 system},
  author={Efimchenko, VS and Antonov, VE and Barkalov, OI and Klyamkin, SN and Tkacz, M},
  journal={High Pressure Research},
  volume={29},
  number={2},
  pages={250--253},
  year={2009},
  publisher={Taylor \& Francis}
}

@article{Belosudov2016a,
  title={Hydrogen hydrates: Equation of state and self-preservation effect},
  author={Belosudov, Rodion V and Bozhko, Yulia Yu and Zhdanov, Ravil K and Subbotin, Oleg S and Kawazoe, Yoshiyuki and Belosludov, Vladimir R},
  journal={Fluid Phase Equilibria},
  volume={413},
  pages={220--228},
  year={2016},
  publisher={Elsevier}
}

@article{Dyadin1999b,
  title={Clathrate formation in water-noble gas (hydrogen) systems at high pressures},
  author={Dyadin, Yu A and Larionov, EG and Aladko, E Ya and Manakov, A Yu and Zhurko, FV and Mikina, TV and Komarov, V Yu and Grachev, EV},
  journal={Journal of Structural Chemistry},
  volume={40},
  pages={790--795},
  year={1999},
  publisher={Springer}
}

@article{Grim2012a,
  title={Synthesis and characterization of sI clathrate hydrates containing hydrogen},
  author={Grim, R Gary and Kerkar, Prasad B and Shebowich, Michele and Arias, Melissa and Sloan, E Dendy and Koh, Carolyn A and Sum, Amadeu K},
  journal={The Journal of Physical Chemistry C},
  volume={116},
  number={34},
  pages={18557--18563},
  year={2012},
  publisher={ACS Publications}
}

@article{Katsumasa2007a,
  title={On the thermodynamic stability of hydrogen clathrate hydrates},
  author={Katsumasa, Keisuke and Koga, Kenichiro and Tanaka, Hideki},
  journal={The Journal of chemical physics},
  volume={127},
  number={4},
  year={2007},
  publisher={AIP Publishing}
}

@article{Papadimitriou2016a,
  title={The effect of lattice constant on the storage capacity of hydrogen hydrates: A Monte Carlo study},
  author={Papadimitriou, Nikolaos I and Tsimpanogiannis, Ioannis N and Economou, Ioannis G and Stubos, Athanassios K},
  journal={Molecular Physics},
  volume={114},
  number={18},
  pages={2664--2671},
  year={2016},
  publisher={Taylor \& Francis}
}

@article{Liu2017a,
  title={Ab initio study of the molecular hydrogen occupancy in pure H2 and binary H2-THF clathrate hydrates},
  author={Liu, Jinxiang and Hou, Jian and Xu, Jiafang and Liu, Haiying and Chen, Gang and Zhang, Jun},
  journal={International journal of hydrogen energy},
  volume={42},
  number={27},
  pages={17136--17143},
  year={2017},
  publisher={Elsevier}
}

@article{Belosludov2009a,
  title={Accurate description of phase diagram of clathrate hydrates at the molecular level},
  author={Belosludov, Rodion V and Subbotin, Oleg S and Mizuseki, Hiroshi and Kawazoe, Yoshiyuki and Belosludov, Vladimir R},
  journal={The Journal of chemical physics},
  volume={131},
  number={24},
  year={2009},
  publisher={AIP Publishing}
}

@article{Papadimitriou2008a,
  title={Evaluation of the hydrogen-storage capacity of pure H2 and binary H2-THF hydrates with Monte Carlo simulations},
  author={Papadimitriou, NI and Tsimpanogiannis, IN and Papaioannou, A Th and Stubos, AK},
  journal={The Journal of Physical Chemistry C},
  volume={112},
  number={27},
  pages={10294--10302},
  year={2008},
  publisher={ACS Publications}
}

@article{Papadimitriou2008b,
  title={Hydrogen storage in sH hydrates: A Monte Carlo study},
  author={Papadimitriou, NI and Tsimpanogiannis, IN and Peters, CJ and Papaioannou, A Th and Stubos, AK},
  journal={The Journal of Physical Chemistry B},
  volume={112},
  number={45},
  pages={14206--14211},
  year={2008},
  publisher={ACS Publications}
}

@article{C5CP07202F,
author ="Conde, M. M. and Torré, J. P. and Miqueu, C.",
title  ="Revisiting the thermodynamic modelling of type I gas–hydroquinone clathrates",
journal  ="Phys. Chem. Chem. Phys.",
year  ="2016",
volume  ="18",
issue  ="15",
pages  ="10018-10027",
}

@article{Chun2008a,
  title={Molecular simulation of cage occupancy and selectivity of binary THF--H2 sII hydrate},
  author={Chun, Dong-Hyuk and Lee, Tai-Yong},
  journal={Molecular Simulation},
  volume={34},
  number={9},
  pages={837--844},
  year={2008},
  publisher={Taylor \& Francis}
}

@article{Papadimitriou2017a,
  title={Storage of H2 in clathrate hydrates: Evaluation of different force-fields used in Monte Carlo simulations},
  author={Papadimitriou, Nikolaos I and Tsimpanogiannis, Ioannis N and Economou, Ioannis G and Stubos, Athanassios K},
  journal={Molecular Physics},
  volume={115},
  number={9-12},
  pages={1274--1285},
  year={2017},
  publisher={Taylor \& Francis}
}

@article{Tanaka2018a,
  title={On the thermodynamic stability of clathrate hydrates VI: complete phase diagram},
  author={Tanaka, Hideki and Yagasaki, Takuma and Matsumoto, Masakazu},
  journal={The Journal of Physical Chemistry B},
  volume={122},
  number={1},
  pages={297--308},
  year={2018},
  publisher={ACS Publications}
}

@article{mi2024molecular,
  title={Molecular insights into the microscopic behavior of {CO}$_2$ hydrates in oceanic sediments: Implications for carbon sequestration},
  author={Mi, Fengyi and Li, Wei and Pang, Jiangtao and Moultos, Othonas A and Ning, Fulong and Vlugt, Thijs JH},
  journal={The Journal of Physical Chemistry C},
  volume={128},
  number={43},
  pages={18588--18597},
  year={2024},
  publisher={ACS Publications}
}

@article{brumby2016cage,
  title={Cage occupancy of methane hydrates from Gibbs ensemble Monte Carlo simulations},
  author={Brumby, Paul E and Yuhara, Daisuke and Wu, David T and Sum, Amadeu K and Yasuoka, Kenji},
  journal={Fluid Phase Equilibria},
  volume={413},
  pages={242--248},
  year={2016},
  publisher={Elsevier}
}

@article{Zhang2024a,
  title={Probing the pathway of H2-THF and H2-DIOX sII hydrates formation: Implication on hydrate-based H2 storage},
  author={Zhang, Jibao and Li, Yan and Rao, Yizhi and Li, Yang and He, Tianbiao and Linga, Praveen and Wang, Xiaolin and Chen, Qian and Yin, Zhenyuan},
  journal={Applied Energy},
  volume={376},
  pages={124289},
  year={2024},
  publisher={Elsevier}
}

@article{Zhang2023b,
  title={How THF tunes the kinetics of H2--THF hydrates? A kinetic study with morphology and calorimetric analysis},
  author={Zhang, Jibao and Li, Yan and Yin, Zhenyuan and Zheng, Xiang Yuan and Linga, Praveen},
  journal={Industrial \& Engineering Chemistry Research},
  volume={62},
  number={51},
  pages={21918--21932},
  year={2023},
  publisher={ACS Publications}
}

@article{Zhang2023c,
  title={Coupling amino acid L-Val with THF for superior hydrogen hydrate kinetics: Implication for hydrate-based hydrogen storage},
  author={Zhang, Jibao and Li, Yan and Yin, Zhenyuan and Linga, Praveen and He, Tianbiao and Zheng, Xiang Yuan},
  journal={Chemical Engineering Journal},
  volume={467},
  pages={143459},
  year={2023},
  publisher={Elsevier}
}

@article{Borrero2025a,
  title={Three-Phase Equilibria of CO2 Hydrate from Computer Simulation in the Presence of NaCl},
  year={2025},
  author={Borrero, A and D{\'\i}az-Acosta, A and Blazquez, S and Zer{\'o}n, IM and Algaba, J and Conde, MM and Blas, FJ},
  journal={Energy \& Fuels},
  publisher={ACS Publications}
}

\end{document}